\documentclass[fleqn,usenatbib,useAMS]{mnras}

\usepackage{graphicx}             % Including figure files
\usepackage{amsmath}              % Advanced maths commands
\usepackage{upgreek}              % uptau
\usepackage{amssymb}              % Extra maths symbols
\usepackage{bm}                   % Bold maths symbols, including upright Greek
\usepackage{CJKutf8}              % Chinese name
\usepackage{comment}              % Comment big block
\usepackage{tabularx}             % Newer table package
\usepackage{threeparttable}       % Table notes smart width
\usepackage{booktabs}             % Better \hline
\usepackage{soul}                 % better underline
\usepackage{ulem}                 % better strikethrough
\usepackage{multirow}             % to enable multirow cells in tables
\usepackage[T1]{fontenc}          % for special char in Drazkowska2014
\usepackage{makecell}
\usepackage[dvipsnames]{xcolor}   % dye text or math symbols (uncomment for MNRAS)
\usepackage{microtype}            % to disable ligatures -- don't like them
\DisableLigatures[f]{encoding = *, family = *} %% <- only disables f-ligatures
\emergencystretch=\maxdimen       % to ease underfull/overfull lines.
\newcommand{\fracp}[2]{\frac{\partial{#1}}{\partial{#2}}}

\def\approxprop{\def\p{
    \setbox0=\vbox{\hbox{$\propto$}}
    \ht0=0.6ex \box0 }
  \def\s{\vbox{\hbox{$\sim$}}}
  \mathrel{\raisebox{0.7ex}{\mbox{$\underset{\s}{\p}$}}}
}
\usepackage{ae,aecompl}
\usepackage{multicol}        % Multi-column entries in tables
\usepackage{newtxtext,newtxmath}

\title[Dust Growth near Truncation Radius]{Dust Accumulation near the Magnetospheric Truncation of Protoplanetary Discs around T Tauri Stars}

\author[R. Li, Y. -X. Chen, and D. N. C. Lin]{
Rixin Li$^{1}$
\begin{CJK*}{UTF8}{gbsn}
  (李日新)
\end{CJK*} \thanks{Contact e-mail: \href{mailto:rixin.li@cornell.edu}{rixin.li@cornell.edu}},
Yi-Xian Chen$^{2,3}$
\begin{CJK*}{UTF8}{gbsn}
  (陈逸贤)
\end{CJK*}
, and Douglas N. C. Lin$^{4,5}$
\begin{CJK*}{UTF8}{gbsn}
  (林潮)
\end{CJK*}
\\
$^{1}$Center for Astrophysics and Planetary Science, Department of Astronomy, Cornell University, Ithaca, NY 14853, USA \\
$^{2}$Department of Astrophysical Sciences, Princeton University, Princeton, NJ 08544, USA \\
$^{3}$Department of Physics, Tsinghua University, Beijing, 100084, People’s Republic of China \\
$^{4}$Department of Astronomy, University of California, Santa Cruz, CA 95064, USA \\
$^{5}$Institute for Advanced Studies, Tsinghua University, Beijing, 100084, People’s Republic of China
}

\date{Accepted 2021 Decemeber 14; in original form 2021 September 24}
\pubyear{2021}
\begin{document}
\label{firstpage}
\pagerange{\pageref{firstpage}--\pageref{lastpage}}
\maketitle

%%%%%%%%%%%%%%%%%%%%%%%%%%%%%%%%%%%%%%%%%%%%%%%%%%%%%%%%%%%%%%%%%%%%%%%%%%%%%%%%
\begin{abstract}
  The prevalence of short-period super-Earths that are independent of host metallicity challenges the theoretical construction of their origin.
  We propose that dust trapping in the global pressure bump induced by magnetospheric truncation in evolved protoplanetary discs (PPDs) around T Tauri stars offers a promising formation mechanism for super-Earths, where the host metallicity is already established. 
  To better understand this planet forming scenario, we construct a toy inner disc model and focus on the evolution of dust trapped in the bump, taking into account the supply from drifting pebbles and loss due to funnel flows.  We develop an implicit coagulation-fragmentation code, \texttt{Rubble}, and perform a suite of simulations to evolve the local dust size distributions.  Our study for the first time considers dust feedback effect on turbulent diffusion in this kind of model.  
  We report that efficient dust growth and significant accumulation of dust mass is possible in less turbulent disc with sturdier solids and with faster external supply, laying out a solid foundation for further growth towards planetesimals and planetary embryos.  We further find that, depending on the dominant process, solid mass may predominantly accumulate in cm-sized grains or particles in runaway growth, indicating different ways of forming planetesimals.  Furthermore, these various outcomes show different efficiencies in saving dust from funnel flows, suggesting that they may be distinguishable by constraining the opacity of funnel flows.  Also, these diverse dust behaviours may help explain the observed dipper stars and rapidly varying shadows in PPDs.
\end{abstract}

% MNRAS

%%%%%%%%%%%%%%%%%%%%%%%%%%%%%%%%%%%%%%%%%%%%%%%%%%%%%%%%%%%%%%%%%%%%%%%%%%%%%%%%
\section{Introduction}
\label{sec:intro}

% 1. General background
% 1.1 super-Earths are common and are insensitive to stellar metallicity
Radial velocity surveys \citep[e.g., ][]{Howard2010, Mayor2011} and transit surveys (e.g., the \textit{Kepler} and \textit{TESS} missions; \citealt{Batalha2013, Fressin2013, Dressing2015, Guerrero2021}, etc.) have found that the most abundant type of planets around solar type stars (including early M stars) are super-Earths ($1-4 R_{\oplus}$, also known as sub-Neptunes or Kepler planets).  Contrary to those of gas giants, the sizes/masses and occurrence rates of super-Earths are generally independent of host metallicity \citep[][see also arguments for a weak dependency in \citealt{ZhuWei2019}]{Buchhave2014, Winn2017, Wu2019, Kutra2021}.
% 1.2 super-Earths are often in tight orbits
Moreover, super-Earths are often found in short-period orbits ($<100$ days) and in multiple systems with compact and coplanar orbits \citep[][etc.]{Fang2013, Fabrycky2014, Dressing2015}.

% 1.3 theoretical challenges and previous work
The tight orbital configurations of super-Earths, the lack of their analogues in the Solar System, and particularly the insensitivity to host metallicity challenges the conventional planet formation scenario.  Previous studies have shown that it might be unlikely for super-Earths to form in situ with local materials and still remain in the close-in orbits since the required protoplanetary disc (PPD) profile is too massive to not alter their orbits \citep[see review by][and references therein]{Morbidelli2016}.

\begin{figure*}%[t]
  \centering
  \includegraphics[width=\linewidth]{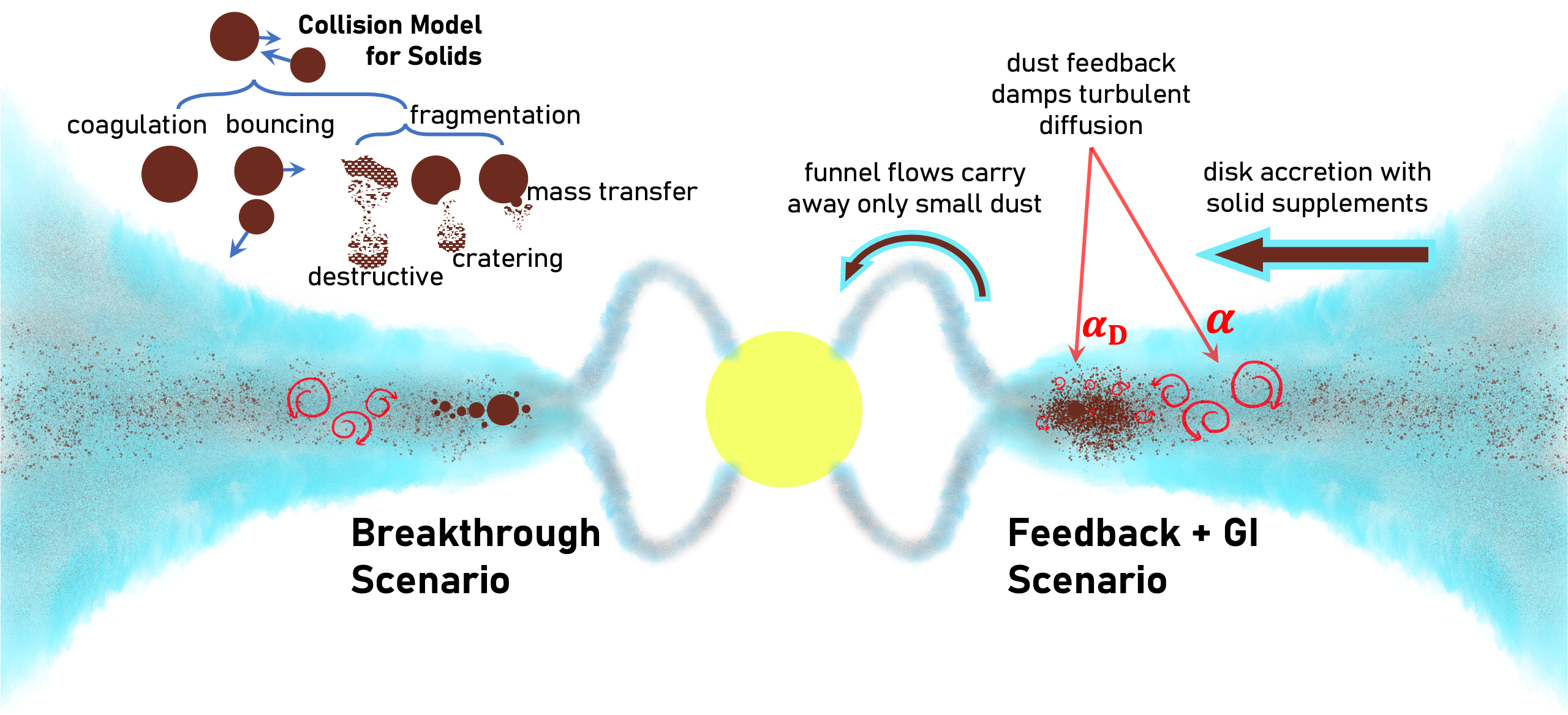}
  \caption{Schematic illustration of dust evolution in the global pressure maximum induced by the magnetospheric truncation, assuming a relatively low accretion rate such that gas temperature allows dust survival.  This work reports two possible scenarios that lead to efficient accumulation of solid materials in the pressure maximum: (a) \textit{Breakthrough Scenario}: weaker turbulent diffusion and supplied large particles break through the fragmentation barrier via modulated mass transfer and produce objects beyond 100 m that sweep-up pebbles; (b) \textit{Feedback + GI Scenario}: dust around cm in size becomes a mass reservoir for fragmented supplied large particles, where the damped turbulent diffusion reduces the amount of pulverized dust carried away by funnel flows.  Such a mass reservoir eventually becomes unstable and produces planetesimals via the gravitational instability.  \label{fig:scenarios}}
\end{figure*}

% 1.3 - previous work on migration scenario
Alternatively, super-Earths may form out of materials from the external disc.  The remaining key question is then when and where the growth takes place.  It is possible that planetary embryos form in the external disc (i.e., at a few au) and then migrate into the current orbital configurations \citep[e.g., ][]{Terquem2007, Ida2010, Kley2012, Cossou2014}.  These embryos are expected to be trapped in a chain of mean motion resonances but few observational matches are found \citep{Mills2016}, suggesting that such resonant systems, if ever populated, would be destabilized afterward \citep[e.g.,][]{Pu2015, Chatterjee2015}.  
Moreover, this scenario holds implicit assumptions that embryos form efficiently, grow via pebble accretion, migrate inward, and stop at the close-in orbits before disc dissipation \citep[e.g.,][]{Lambrechts2014, Bitsch2015, Johansen2017, Lambrechts2019, Izidoro2021}, all of which introduce uncertainties and merit further investigations.  % RL: add examples that migration timescale exceeds the disc lifetime, Kepler 444?

% 1.3 - previous work on in situ formation with drifting pebbles
An alternate scenario invokes accumulation of solids drifted from the external disc followed by in situ growth from dust to planetesimals or planetary embryos.  Such a scenario relies on rapid inward drift of dust particles due to gas drag \citep{Weidenschilling1977} and a dust trapping mechanism.  This scenario is thus sensitive to the pebble delivery efficiency through the disc, which may be hindered by disc substructures \citep[e.g.,][]{Andrews2018, Dullemond2018, Pinilla2018, Pinilla2020}.  That said, ALMA disc surveys indicate that smooth compact discs may be common \citep{Long2019}, suggesting this scenario is a promising formation pathway.  It is also worth noting that some of these smooth discs may have hypothetical underlying substructures hitherto unresolved \citep{Jennings2021}.

% 1.3 - previous work on the IOPF and its limitations
Pressure bumps are nature traps for pebbles \citep{Pinilla2012a, Pinilla2017}.  \citet{Chatterjee2014} hypothesized the inside-out planet formation (IOPF) theory where pebbles are trapped in the local pressure maximum at the dead zone inner boundary (DZIB), which eventually lead to sequential planet formation \citep[see also][etc.]{Chatterjee2015, HuXiao2016, HuXiao2018}.  However, recent inner disc models suggest that the DZIB-induced pressure maximum, if exists, may locate far out ($\gtrsim 0.3$ au) \citep{Jankovic2021b, Jankovic2021}, potentially inconsistent with the close-in orbits of super-Earths.  Moreover, the IOPF theory so far seems to neglect the important observational finding that super-Earths are insensitive to their host metallicity.

% 1.4 bring up possible solution and observational motivations
Instead of the DZIB, a natural global pressure maximum exists at the inner edge of PPDs -- the magnetospheric truncation radius $R_{\rm T}$ (\citealt{Koenigl1991}; see also reviews by \citealt{Dullemond2010}, \citealt{Lai2014}, and \citealt{Hartmann2016}) -- and provides a higher dust trapping efficiency than local pressure maxima.  The gas temperature around $R_{\rm T}$ is too high for refractory grains to survive \textit{unless} in evolved PPDs (e.g., late Class II discs and afterward), where the relatively low accretion rates result in an expanded $R_{\rm T}$ and consequently a cooler temperature.  Therefore, the global pressure maximum by then serves as an ideal dust trap and a promising site for planet formation.

Such a scenario directly solves the metallicity conundrum.  At the late stage of disc evolution, the central protostars are expected to have acquired nearly all their asymptotic masses and heavy elements, which disconnects their metallicity from planets formed at this stage.  Moreover, the truncation radius $R_{\rm T}$ is likely to still be inside $0.3$ au when the gas temperature becomes survivable for refractory grains, making planets formed near $R_{\rm T}$ more likely to be consistent with observed close-in super-Earths.

Furthermore, although the detailed inner disc structures are poorly understood \citep{Dullemond2010}, dust near the inner disc edge has long been proposed to explain the dippers -- a class of young stellar objects (YSOs) with large transient drops in flux -- by occultation \citep[e.g.,][]{Bouvier1999, Cody2014, Bodman2017, Hedges2018, Roggero2021}.  Some dippers show correlations between dimming patterns and stellar rotation, suggesting that dust is around the corotation radius $R_{\rm Co}$ \citep{Stauffer2017}, which is closely related to and may be roughly equal to $R_{\rm T}$ \citep{Long2005, Bouvier2007}.

Motivated by the encouraging and long-standing observational evidence of dust features near the inner edge of PPDs, we investigate the scenario of planet formation near $R_{\rm T}$ by studying the local evolution of solids in the global pressure maximum.  In this work, we are particularly interested in the evolution of dust size distribution regulated by accretion and the resulting dust growth and accumulation (illustrated in Figure \ref{fig:scenarios}).

The paper is organized as follows.  In Section \ref{sec:inner_disc_models}, we explore and construct the inner disc model that allow dust survival in the pressure maximum near $R_{\rm T}$.  Section \ref{sec:removal_processes} then describes and quantifies the size-dependent removal processes for dust grains via magnetic fields and funnel flows.  Section \ref{sec:coag_model} details our numerical tool for modelling dust size distribution evolution.  We propose to incorporate dust feedback effect in Section \ref{subsubsec:feedback} and dynamic mass exchange with accretion flows in Section \ref{subsec:dust_evo}.  Section \ref{subsec:setup} lays out the parameter space covered by our simulations.  We analyze and present dust evolution results in Section \ref{sec:results}, followed by discussions on implications and limitations in Section \ref{sec:final}.

%%%%%%%%%%%%%%%%%%%%%%%%%%%%%%%%%%%%%%%%%%%%%%%%%%%%%%%%%%%%%%%%%%%%%%%%%%%%%%%%
\section{Inner Disc Models}
\label{sec:inner_disc_models}

In this section, we quantify disc conditions near the magnetospheric truncation radius and identify the relevant parameter space that allows dust survival.  Through this study, we focus on a solar-type pre-main-sequence star with mass $M_\star=M_{\odot}$.  Based on the stellar evolution model in \citet{Baraffe2015},
\footnote{\href{http://perso.ens-lyon.fr/isabelle.baraffe/BHAC15dir/BHAC15\_tracks+structure}{http://perso.ens-lyon.fr/isabelle.baraffe/BHAC15dir/BHAC15\_tracks+structure}}
we consider a typical range of stellar radius $R_{\star} \in [R_{\odot}, 2R_{\odot}]$.

%%%%%%%%%%%%%%%%%%%%%%%%%%%%%%%%%%%%%%%%%%%%%%%%%%
\subsection{Magnetospheric Truncation Radius}
\label{subsec:R_T}

T Tauri stars provides sufficient magnetic torques on disc gas and clear out a cavity to the magnetospheric truncation radius \citep{Konigl2011}
\begin{equation}
  R_{\rm T} \simeq \left(\frac{B_\star^4 R_\star^{5}}{G M_\star \dot{M}^2} \right)^{1/7} R_\star, \label{eq:R_T}
\end{equation}
where $\dot{M}$ and $B_\star$ are the accretion rate and the strength of the stellar dipole magnetic field.  This truncation radius increases with decreasing accretion rate, making it possible for the truncation induced pressure maximum to capture dust before sublimation when $\dot{M}$ becomes relatively low.

We further assume that the corotation radius
\begin{equation}
  R_{\rm Co} = \left[ \frac{GM_\star}{(2\pi/P_\star)^2} \right]^{1/3}
\end{equation}
converges with the truncation radius
\footnote{In early stages of stellar/disc evolution, the angular momentum transport due to disc-magnetosphere interactions locks the protostar into the rotational equilibrium state, where $R_{\rm Co}/R_{\rm T} \sim 1.4$ \citep{Long2005, Romanova2008}.  Our scenario concerns later stages, where $R_{\rm T}$ tends to expand and converges with $R_{\rm Co}$ as $\dot{M}$ gradually decreases.}
, where $P_\star$ is the stellar rotation period.  The rotation periods of T Tauri stars peak at $\sim 8$ days \citep[e.g.,][]{Attridge1992, Herbst2002, Bouvier2007b, Leechiang2017}.  We thus adopt $P_\star = 8$ days in this work and determine the accretion rate with
\begin{equation}
  \dot{M} = \frac{B_{\star}^{2}R_{\star}^{6}}{(GM_{\star})^{5/3}} \left(\frac{2\pi}{P_\star}\right)^{7/3},
  \label{eq:mdotstar_corot}
\end{equation}
which is the fundamental quantity for constructing disc profiles in the following section. 

%%%%%%%%%%%%%%%%%%%%%%%%%%%%%%%%%%%%%%%%%%%%%%%%%%
\subsection{Radiative Disc Model}
\label{subsec:rad_disc_profile}

To study the very inner disc region ($\lesssim 1$ au), we adopt the radiative disc model from \citet{Ali-Dib2020} where viscous heating dominates over stellar irradiation \citep{GaraudLin2007} and heat is transported vertically through radiative diffusion.  The radial profiles for temperature and gas density are
\begin{align}
  T_{\rm rad}=& 373 \mathrm{~K}\ \ r_{\mathrm{au}}^{-9 / 10} \alpha_{-2}^{-1 / 5} \dot{M}_{-7.5}^{2 / 5} m_\star^{3/10} \kappa_{\rm cmg} ^{1/5}, \label{eq:T_rad} \\[6pt]
  \rho_{\rm rad}=& 1.7 \times 10^{-10}\ \text{g cm}^{-3} r_{\mathrm{au}}^{-33 / 20} \alpha_{-2}^{-7 / 10} \dot{M}_{-7.5}^{2 / 5} m_\star ^{11/20} \kappa_{\rm cmg} ^{-3/10}
\end{align}
where $\alpha_{-2} = \alpha / 0.01$ and $\alpha$ is the Shakura-Sunyaev parameter, $\dot{M}_{-7.5} = 10^{-7.5}M_\odot \text{yr}^{-1}$, $m_\star = M_\star/M_\odot$, $\kappa_{\rm cmg} = \kappa/1{\rm \ cm^2 \ g^{-1}}$ and $\kappa$ is the opacity, and $r_{\rm au} = R/(1{\rm \ au})$ and $R$ is the disc radius.  The gas surface density in this model is $\Sigma_{\rm g} = 2 \rho_{\rm rad} H$, where $H$ is the gas scale height.  Our calculations assume a steady gas disc with a constant opacity for simplicity.  We discuss the self-consistency of this assumption in Section \ref{sec:final}.

%%%%%%%%%%%%%%%%%%%%%%%%%%%%%%%%%%%%%%%%%%%%%%%%%%
\subsection{Viable Accretion Rates}
\label{subsec:viable_Mdot}

We now calculate the disc conditions that make dust trapping and retention permissible.  First, we assume that the dust accumulation site (i.e., the truncation induced global pressure maximum) is slightly outside $R_{\rm T}$ and parameterize its radial location with
\begin{equation}
  R_{\rm accu} = f_{\rm out} R_{\rm T}, \label{eq:R_accu}
\end{equation}
where $f_{\rm out}$ is fixed to $1.25$ throughout this work.  We adopt such a simple scaling because the radiative disc model does not include a truncated cavity.  We defer self-consistent modelling of truncated disc profiles to future work. 

The gas temperature $T_{\rm rad}$ is then the key quantity to constrain relevant disc conditions.  First, disc truncation is only attainable when $T_{\rm rad}$ at $R_{\rm T}$ is above $\sim 10^3$ K \citep{Umebayashi1988, DeschTurner2015}.  Such a temperature is required for alkali metals to sublimate and make the gas disc ionized enough so that disc-magnetosphere coupling is sufficient to launch funnels flows and create a cavity.  Second, $T_{\rm rad}$ in the global pressure maximum should be lower than the dust sublimation temperature, $\sim$2000 K \citep[][etc.]{Kobayashi2011}.  Otherwise, all solids become gas before reaching the trapping site and subsequently become metal pollution upon accretion.

\begin{figure}
  \centering
  \includegraphics[width=\linewidth]{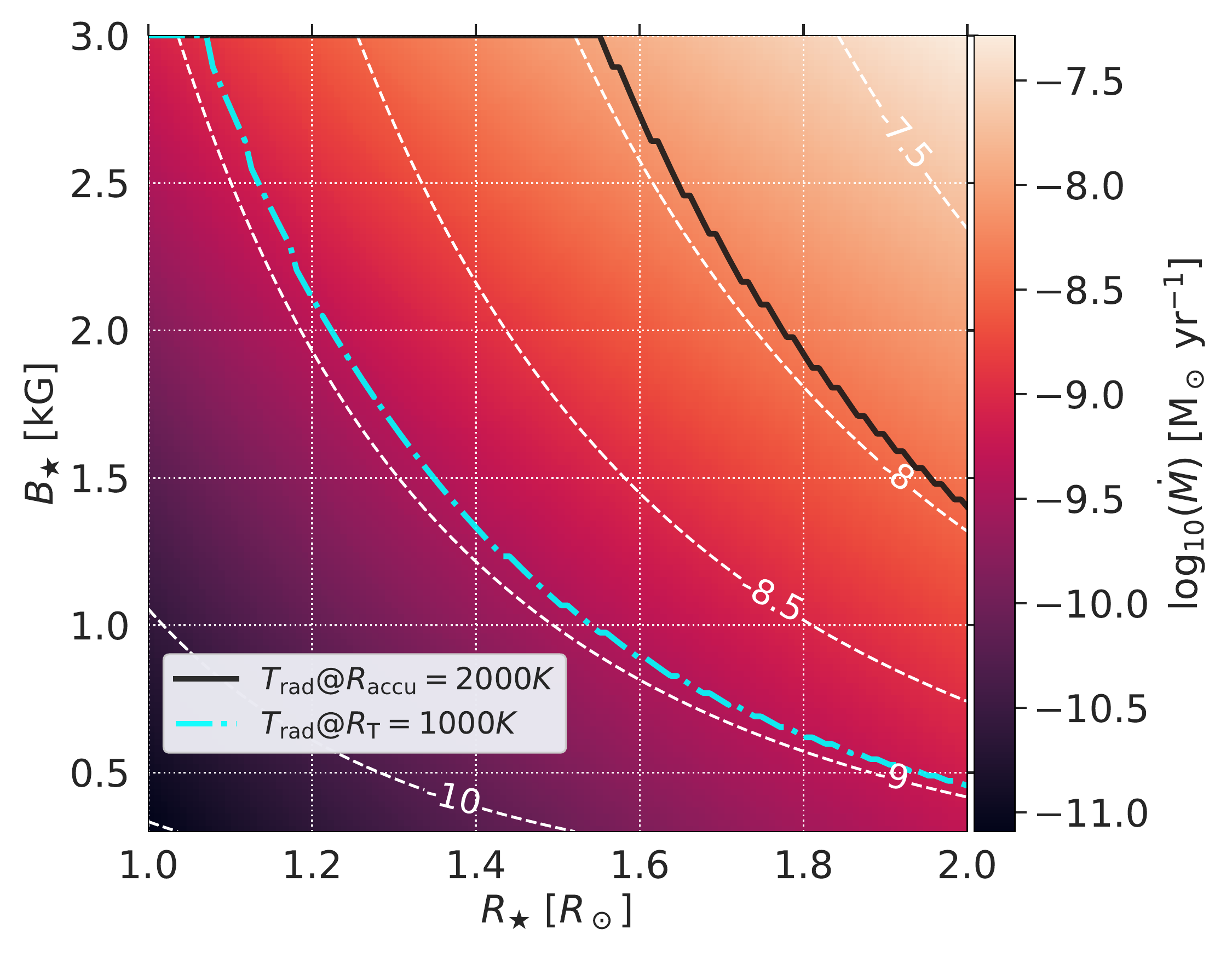}
  \caption{The accretion rate (see Equation \ref{eq:mdotstar_corot}) in the plane of stellar radius and the strength of the stellar magnetic field, assuming $M_\star=M_\odot$, $P_\star=8$ days, $\alpha=0.01$.  The \textit{black solid} curve and the \textit{blue dash dotted} curve sandwich the area where the disc temperature $T_{\rm rad}$ at $R_{\rm T}$ is above $10^3$ K (such that disc is truncated) and $T_{\rm rad}$ at $R_{\rm accu}$ is below $2000$ K (such that dust do not sublimate), allowing dust trapping and survival.
  \label{fig:Mdot_map}}
\end{figure}

Figure \ref{fig:Mdot_map} shows the accretion rate for a typical range of stellar radius and stellar magnetic field strength \citep{Johns-Krull2007, Yang2008, Yang2011} with the assumption $\alpha=0.01$.  We use the temperature requirements as a proxy to identify viable combinations of stellar properties and accretion rates that lead to $T_{\rm rad}(R_{\rm T}) > 1000$ K and $T_{\rm rad}(R_{\rm accu}) < 2000$ K.  These accretion rates lie between $\sim 10^{-9}$ -- $\sim 10^{-8} M_\odot/$ yr, consistent with the late stages of disc evolution.

%%%%%%%%%%%%%%%%%%%%%%%%%%%%%%%%%%%%%%%%%%%%%%%%%%
\begin{figure*}
  \centering
  \includegraphics[width=0.495\linewidth]{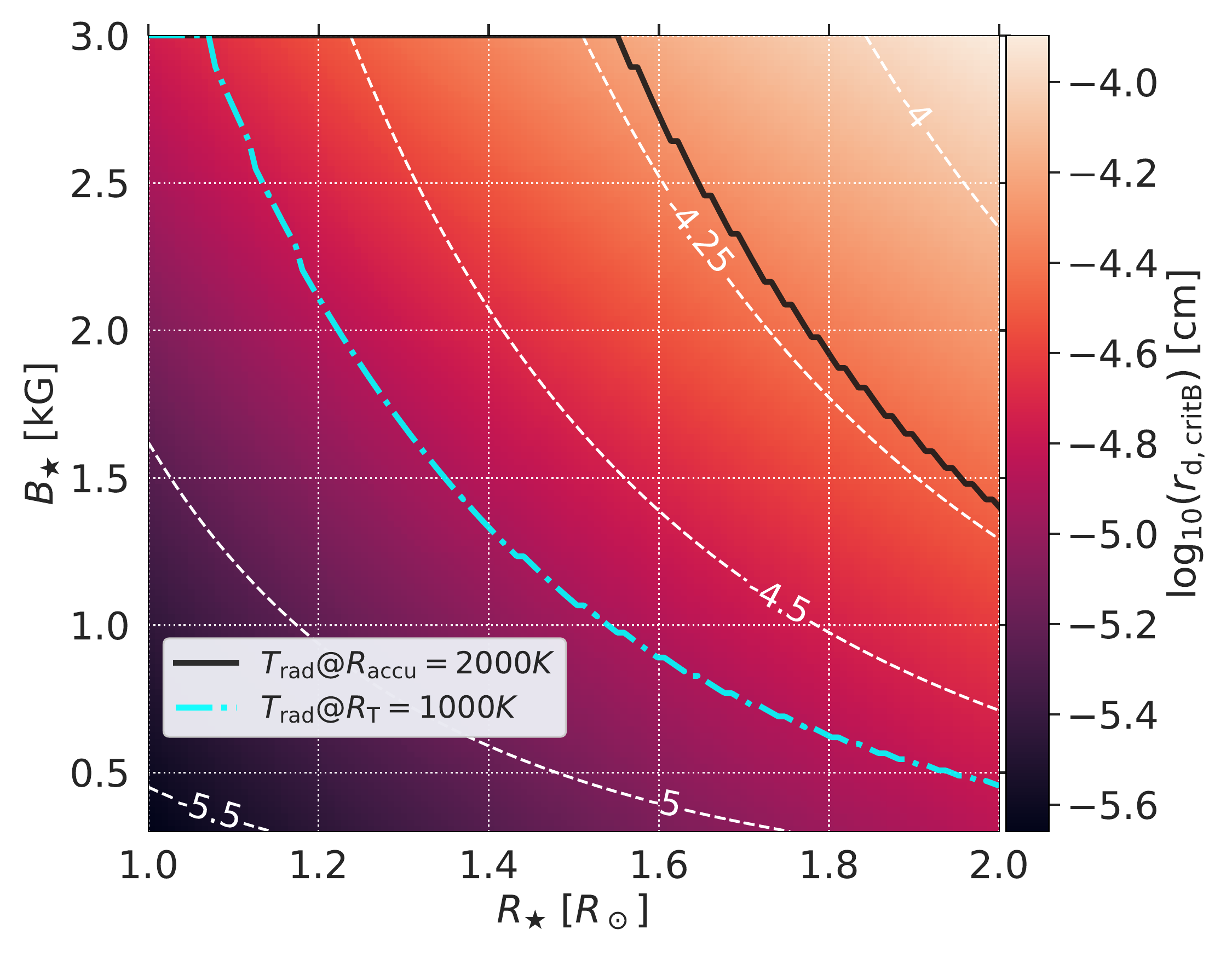}
  \includegraphics[width=0.495\linewidth]{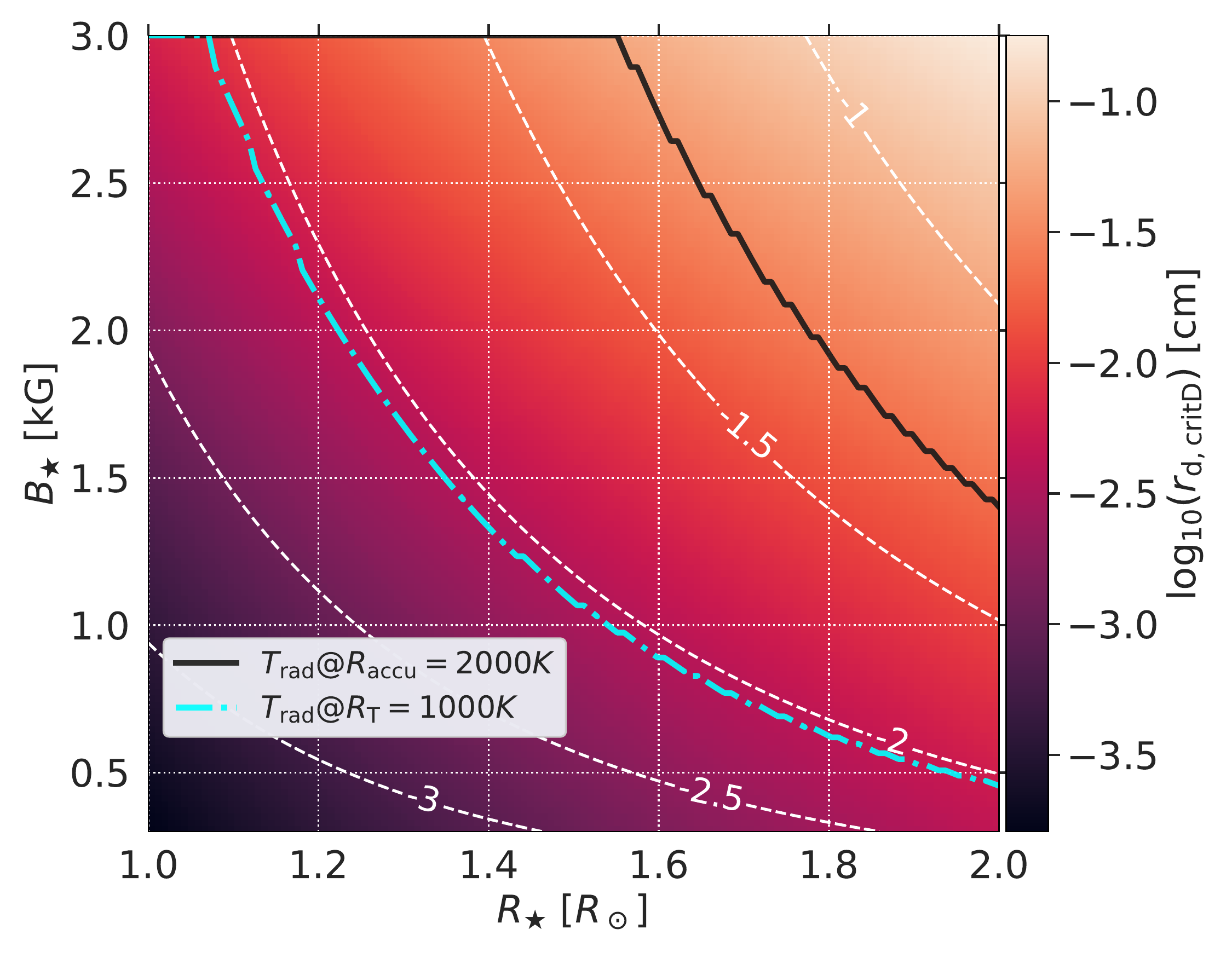}
  \caption{Similar to Figure \ref{fig:Mdot_map} but for the critical particle size below which dust would be removed by Lorentz force from magnetic field (\textit{left}; see Equation \ref{eq:r_d_critB}) and by drag force from funnel flows (\textit{right}; see Equation \ref{eq:r_d_critD}).
  \label{fig:r_crit_map}}
\end{figure*}

%%%%%%%%%%%%%%%%%%%%%%%%%%%%%%%%%%%%%%%%%%%%%%%%%%%%%%%%%%%%%%%%%%%%%%%%%%%%%%%%
\section{Dynamical Dust Removal Processes} 
\label{sec:removal_processes}

The fate of dust trapped at the global pressure maximum $R_{\rm accu}$ is subject to grain size.  While larger solids may be retained within the bump, smaller particles experience dynamical removal processes.  In this section, we focus on such processes resulted from magnetic field and funnel flows.

%%%%%%%%%%%%%%%%%%%%%%%%%%%%%%%%%%%%%%%%%%%%%%%%%%
\subsection{Charged Grains in Magnetic Field}
\label{subsec:charged_grain_mag}

Grains that carry charge may be lifted away by Lorentz force
\begin{equation}
  \bm{F}_{\rm L} = Q_0 \bm{v}_{\rm d} \times \bm{B},
\end{equation}
where $Q_0$ is the grain charge, $\bm{v}_{\rm d}$ is the dust velocity relative to the stellar rotation (assuming filed lines corotate with the central star), $\bm{B}$ is the stellar magnetic field strength at a certain disc radius.
\footnote{The total magnetic field includes the unperturbed stellar magnetic field $\bm{B}$ and the induced field $\bm{B}'$.  However, the magnitude of $\bm{B}'$ is usually negligible in the radial direction when compared to $\bm{B}$ (Lin, Tang, et al., in preparation).  Thus, $\bm{B}'$ contributes little to the vertical component of Lorentz force on dust and is neglected.}
For simplicity, we are interested in the vertical component of $\bm{F}_{\rm L}$
\begin{equation}
  F_{\mathrm{L}, z} = \left| Q_0 (\Omega_\star - \Omega_{\rm K}) R B_r \right|, \label{eq:F_Lz}
\end{equation}
where $\Omega_\star=2\pi/P_\star$ is the angular velocity of stellar rotation, $\Omega_K$ is the Keplerian angular velocity at disc radius $R$, and the radial component of $\bm{B}$ is
\begin{equation}
  B_r = \frac{3 \mathscr{m}_\star z R}{(R^2+z^2)^{5/2}} \simeq \frac{3 \mathscr{m}_\star z}{R^4} %\left(1 - \frac{5z^2}{2R^2} \right),
\end{equation}
where $\mathscr{m}_\star \simeq B_\star R_\star^3$ is the stellar dipole moment and $z$ is the vertical distance to disc midplane.

To estimate the characteristic charge carried by grains $Q_0$, we adopt a disc model with X-ray ionization considered in \citet{Ke2012}.  They found that the timescale for charge loading on grains is very short.  In the equilibrium state, grains are negatively charged to
\begin{equation}
    Q_0 = -5 \frac{4 \pi \epsilon_0 r_{\rm d} E_e}{e}, \label{eq:Q0}
\end{equation}
where $e$ is the unit electron charge, $\epsilon_0$ is the vacuum permittivity, $r_{\rm d}$ is the grain radius, $E_e = (3/2)k_{\rm B} T_e$ is the energy of electrons, and $T_e$ is approximated by $T_{\rm rad}$ in this work.

We are now able to estimate the critical particle size below which dust can be lifted away at $R_{\rm accu}$.  Equating $F_{\mathrm{L},z}$ to the vertical stellar gravity gives
\begin{align}
    &\left| Q_0 (\Omega_\star - \Omega_K ) R_{\rm accu} B_{\rm r} \right| \simeq m_{\rm d} \Omega_K^2 z,  \\
    \Rightarrow\ &r_\mathrm{d, critB} \simeq \left| 45 \frac{\epsilon_0 E_e}{e \rho_\bullet} \frac{\Omega_\star - \Omega_{\rm K}}{\Omega_{\rm K}^2} \frac{\mathscr{m}_\star}{R_{\rm accu}^3} \right|^{1/2}, \label{eq:r_d_critB}
\end{align}
where the grain mass $m_{\rm d}$ is substituted with $4\pi \rho_{\bullet} r_{\rm d}^3/3$, $\rho_\bullet$ is the material density of dust, and $\Omega_{\rm K}$ is the Keplerian frequency at $R_{\rm accu}$ such that $\Omega_\star/\Omega_{\rm K} = f_{\rm out}^{3/2}$.  In this work, we further assume that $\rho_{\bullet} = 2.3$ g/cm$^3$ (i.e., the average material density of silicate grains).  \citet{Birnstiel2011} showed that the dust size distribution varies little when $\rho_{\bullet}$ falls between $1.6$ and $3.0$ g cm$^{-3}$.

Figure \ref{fig:r_crit_map} plots $r_\mathrm{d, critB}$ in the $R_\star$--$B_\star$ space similar to that in Figure \ref{fig:Mdot_map}.  We find that $r_\mathrm{d, critB}$ is overall small ($\ll \mu$m) and is below the size range considered in our dust evolution model (see Section \ref{sec:coag_model}).  We thus neglect the effect of magnetic field on charged grains in our numerical calculations hereafter.

%%%%%%%%%%%%%%%%%%%%%%%%%%%%%%%%%%%%%%%%%%%%%%%%%%
\subsection{Funnel Flows}
\label{subsec:funnel_flows}

Small dust that is well coupled to gas may be carried away by funnel flows via drag force.  We consider funnel flows launched from a disc annulus between $R$ and $R+\textnormal{d}R$ through a transonic surface at height $H_{\rm s}$, where the flow speed is sound speed $c_{\rm s}$.  The steady-state flux of disc gas can be approximated as
\begin{equation} \label{eq:mdotdisc}
    {\dot M} \simeq {\dot \Sigma}_{\rm s} 2 \pi R \textnormal{d}R,
\end{equation}
where ${\dot \Sigma}_{\rm s} \simeq 2 \rho_{\rm s} c_{\rm s}$ is the surface density flux, $\rho_{\rm s}$ is the gas density at the transonic surface, and a factor of $2$ is due to the vertical symmetry (i.e., funnel flows both above and below disc midplane).  The mean free path of gas around $H_{\rm s}$ is
\begin{align}
  \lambda_{\rm mfp} &= \frac{\mu}{\sigma(H_2) \rho_{\rm s}} \simeq \frac{\mu}{\sigma(H_2)} \frac{4 \pi c_{\rm s} R^2 (\textnormal{d}R/R)}{\dot M}. \\
  \begin{split}
    \simeq &1.64\times10^3\ \text{cm}\ \left(\frac{\dot{M}}{3\times10^{-9}\ M_\odot\ \text{yr}^{-1}}\right)^{-1} \\
  &\left(\frac{T}{1000\ \text{K}}\right)^{1/2} \left(\frac{R}{0.1\ \text{au}}\right)^{2} \left(\frac{\textnormal{d}R/R}{0.03}\right).
  \end{split}
\end{align}
where $\mu = 2.34 m_{\rm p}$ is the mean molecular weight in proton masses and $\sigma(H_2) = 2\times 10^{-15}$ cm$^2$ is the cross section of molecular hydrogen.

Given the large $\lambda_{\rm mfp}$, we apply the Epstein Drag law on dust particles around the funnel flow launching points to calculate the drag force
\begin{equation}
    F_{\rm D} \simeq \rho_{\rm s} \pi r_{\rm d}^2 c_{\rm s}^2,
\end{equation}
where a factor of order unity may be added depending on other assumptions.  We now estimate the critical particle size below which dust can be dragged away by balancing the drag force with the vertical stellar gravity
\begin{align}
    \rho_{\rm s}\pi r_{\rm d}^2 c_{\rm s}^2 &\simeq m_{\rm d} \Omega_K^2 z, \\
    \Rightarrow r_{\rm d, critD} &\simeq \frac{3\rho_{\rm s} H}{4 \rho_{\bullet} }, \label{eq:r_d_critD}
\end{align}
where we assume that $H_{\rm s} \sim H$ and apply the substitution $H = c_{\rm s}/\Omega_{\rm K}$.  Since $r_{\rm d, critD} \propto \rho_{\rm s} \propto c_{\rm s}^{-1} \propto T^{-1/2}$, lower gas temperature leads to larger critical dust size. 

Figure \ref{fig:r_crit_map} also presents $r_{\rm d, critD}$ in the $R_\star$--$B_\star$ space similar to that in Figure \ref{fig:Mdot_map}.  We find that $r_\mathrm{d, critD}$ is of the order $0.03$ cm, much larger than the critical size that would be affected by Lorentz force.  These sub-mm grains are within the size range considered in our dust evolution model (see Section \ref{sec:coag_model}) and have the potential to contribute rapid mass loss depending on the size distribution.

To further quantify the relative amount of dust removed by drag forces, we assume that funnel flows only and continuously carry away solids smaller than $r_{\rm d,critD}$ and beyond the launching point ($> H$).  We then adopt a Gaussian profile for the dust vertical distribution
\begin{equation}
  \rho_{\rm d}(z, r_{\rm d}) = \rho_{\rm d, 0} (r_{\rm d}) \exp\left(-\frac{z^2}{2H_{\rm d}^2 (r_{\rm d})}\right),
\end{equation}
where $\rho_{\rm d}$ is the dust volume density and $H_{\rm d}$ is the dust scale height determined by grain size and turbulence strength \citep{Youdin2007}
\begin{equation}
  H_{\rm d} (r_{\rm d}) = H \left(1 + \frac{\uptau_{\rm s} (r_{\rm d})}{\alpha} \right)^{-1/2},
\end{equation}
where $\uptau_{\rm s} (r_{\rm d})$ is the dimensionless stopping time \citep[also known as the Stokes number;][]{Birnstiel2010, Youdin2013}:
\begin{eqnarray}\label{eqn:stokes}
  \uptau_{\rm s} (r_{\rm d}) = \left\{
  \begin{aligned}
    & \frac{\pi \rho_{\bullet} r_{\rm d}}{2 \Sigma_{\rm g}} &\mathrm{if}\ r_{\rm d} \leqslant \frac{9\lambda_{\rm mfp}}{4} \ \ \text{(Epstein regime)} \\
    & \frac{2\rho_{\bullet} r_{\rm d}^2}{9 \nu_{\rm mol} \rho_{\rm g}} &\mathrm{if}\ Re<1 \ \  \text{(Stokes regime)} \\
    &\frac{2^{0.6}\rho_{\bullet} r_{\rm d}^{1.6}}{9 \nu_{\rm mol}^{0.6} \rho_{\rm g} u^{0.4}} &\mathrm{if}\ 1 \leqslant Re \leqslant 800  \\ 
    &\frac{6\rho_{\bullet} r_{\rm d}}{\rho_{\rm g} u} &\mathrm{if}\ Re>800 
  \end{aligned}
  \right. ,
\end{eqnarray}
where $\nu_{\rm mol} = 0.5 \bar{u}{\lambda_{\rm mfp}}$ is the gas molecular viscosity, $\bar{u}=\sqrt{\pi/8}c_{\rm s}$ is the mean thermal velocity, $Re=2 r_{\rm d} u / \nu_{\rm mol}$ is the \textit{particle} Reynolds-number, and $u$ denotes the velocity of the dust particle relative to gas.

With the knowledge of dust vertical distribution, the fraction of the dust disc mass beyond the funnel flow launching point ($\sim H$) is thus
\begin{equation} \label{eq:f_H}
    f_H (r_{\rm d}) = \dfrac{2\int_H^\infty \rho_{\rm d }(z, r_{\rm d}) \textnormal{d}z}{\Sigma_{\rm d} (r_{\rm d})} = 1 - \mathrm{erf}\left(\frac{1}{\sqrt{2}}\frac{H}
    {H_{\rm d} (r_{\rm d})} \right),
\end{equation}
where $\Sigma_{\rm d}=\int_{-\infty}^\infty \rho_{\rm d} \textnormal{d}z$ is the dust surface density.  In the next section, we incorporate $f_{\rm H}$ in the dust removal treatment adopted by our numerical model (see \S\ref{subsec:dust_evo}).

%%%%%%%%%%%%%%%%%%%%%%%%%%%%%%%%%%%%%%%%%%%%%%%%%%%%%%%%%%%%%%%%%%%%%%%%%%%%%%%
\section{Dust Evolution Model}
\label{sec:coag_model}

To model the evolution of dust distribution and dust surface density at $R_{\rm accu}$, we develop an implicit coagulation-fragmentation code, \texttt{Rubble}
\footnote{The code is available at \href{https://github.com/astroboylrx/Rubble}{https://github.com/astroboylrx/Rubble}.} \citep{Rubble}.  In this section, we summarize the implementation of our numerical model.  Also, we use $a$ to denote grain size hereafter such that our numerical descriptions can be distinguished from previous analytical estimations.  Appendix \ref{app:tests} demonstrates the robustness of our code via an extensive set of tests.

%%%%%%%%%%%%%%%%%%%%%%%%%%%%%%%%%%%%%%%%%%%%%%%%%%
\subsection{The Base Model}
\label{subsec:base_model}

%%%%%%%%%%%%%%%%%%%%%%%%%%%%%%%%%%%%%%%%%%%%%%%%%%
\subsubsection{The Coagulation-Fragmentation Equation}
\label{subsubsec:smoluchowski}

Our implicit numerical scheme is based on the descriptions in \citet{Birnstiel2010}.  \texttt{Rubble} solves the Smoluchowski equation
\begin{equation}\label{eq:Seq}
    \fracp{}{t} N(m) = \int\int_0^\infty M(m, m', m'') N(m') N(m'') \textnormal{d}m' \textnormal{d}m'',
\end{equation}
where $N(m)\equiv \textnormal{d}N/\textnormal{d}m$ is the vertically integrated dust surface number density in a mass interval, $M(m, m', m'')$ is the coagulation/fragmentation kernel
\begin{equation}\label{eq:kernel}
  \begin{split}
      M&(m, m', m'') = \\
      &\frac{1}{2} K(m', m'')\cdot \delta(m'+m''-m) - K(m', m'')\cdot \delta(m''-m) \\
      &+ \frac{1}{2} L(m', m'')\cdot S(m, m', m'') - L(m', m'')\cdot \delta(m-m''),
  \end{split}
\end{equation}
where $K$ and $L$ are the coagulation and fragmentation kernels, respectively,
\begin{align}
    K(m_1, m_2) &= \Delta u(m_1, m_2) \sigma_{\rm geo}(m_1, m_2) \cdot p_{\rm c}, \\
    L(m_1, m_2) &= \Delta u(m_1, m_2) \sigma_{\rm geo}(m_1, m_2) \cdot p_{\rm f},
    \label{Eqn: KL}
\end{align}
where $\Delta u(m_1, m_2)$ denotes the relative velocity of the two particles, $\sigma_{\rm geo}(m_1, m_2)$ is the geometrical cross section of the collision, and $p_{\rm c}$ and $p_{\rm f}$ are the probabilities for coagulation and fragmentation, respectively (see Section \ref{subsubsec:collisional_outcomes} for their formulae).  In addition, $S$ in Equation \ref{eq:kernel} denotes the distribution of fragments after a complete fragmentation and is described by a power law
\begin{equation}\label{eq:S_power_law}
    N(m)dm \propto m^{-\xi} dm,
\end{equation}
where $\xi = 1.83$ \citep{Brauer2008, Birnstiel2010}.

The total dust surface density $\Sigma_{\rm d}$ can be related to $N(m)$ through the vertically integrated dust surface density distribution per logarithmic bin of grain radius $\sigma(a)$ by 
\begin{equation}
  \Sigma_{\rm d} = \int_0^{\infty} \sigma(a)\ \textnormal{d} \log a,
\end{equation}
where
\begin{equation}
  \sigma(a) = N(m) \cdot 3 m^2 = \frac{dN}{d\log m} 3 m,
\end{equation}
where $dN/d\log m$ is the vertically integrated dust surface number density in a logarithmic mass interval and is the quantity that our implicit code actually evolves.

The Smoluchowski equation is inherently very stiff and the implicit scheme adopted in this work is first-order accurate in time.  The rounding errors also depend on the shape of the particle size distribution.  Therefore, an automatic adaptive time-stepping scheme is employed in \texttt{Rubble} to guarantee the equation-solving meets the desired precision (see also Section \ref{subsec:setup}).

%%%%%%%%%%%%%%%%%%%%%%%%%%%%%%%%%%%%%%%%%%%%%%%%%%
\subsubsection{Relative Velocities between Particles}
\label{subsubsec:dv}

\begin{figure}%[t!]
  \includegraphics[width=\linewidth]{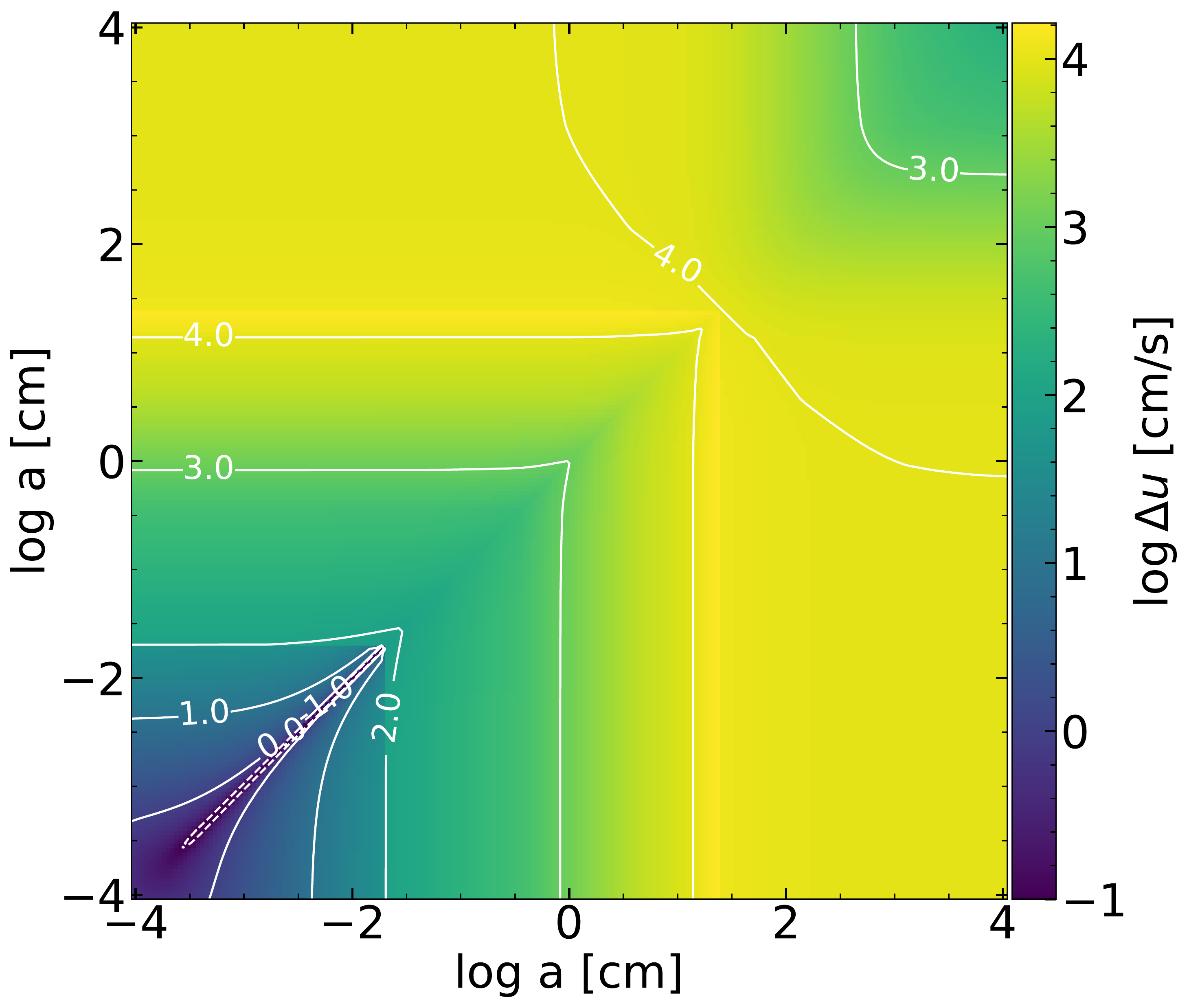}
  \caption{Total relative velocities between dust particles in the initial setup of Model \texttt{A}-series in Section \ref{subsec:fiducial_cases} (see also Section \ref{subsubsec:dv} and Table \ref{tab:fiducial}), including Brownian motions and turbulent relative motions.
  \label{fig:dv_tot}}
\end{figure}

We are interested in solids trapped in the pressure bump at $R_{\rm accu}$, where the temperature is high and the main sources of relative velocities between particles are Brownian motions and gas turbulence \citep{Ormel2007}.  The relative velocities due to differentiated drifting or orbital velocities are thus neglected.  Figure \ref{fig:dv_tot} shows the total relative velocities $\Delta u$ as a function of grain sizes in the initial setup of one of our fiducial models in Section \ref{subsec:fiducial_cases}, where the maximum collision velocity exceeds $10^{4}$ cm s$^{-1}$.

%%%%%%%%%%%%%%%%%%%%%%%%%%%%%%%%%%%%%%%%%%%%%%%%%%
\subsubsection{Collisional Outcomes}
\label{subsubsec:collisional_outcomes}

We take into account three main categories of collisional outcomes, namely coagulation, bouncing, and fragmentation (see Figure \ref{fig:scenarios}).  To statistically determine the outcome of collisions between each pair of grain sizes, we further consider a Maxwellian velocity distribution $\Delta v$, where the root-mean-square velocity is given by the size-specific relative velocity $\Delta u$ \citep[see Section \ref{subsubsec:dv};][]{Windmark2012b}
\begin{equation}
    P(\Delta v | \Delta u) = \sqrt{\frac{54}{\pi}}\frac{\Delta v^2}{\Delta u^3} \exp\left(-\frac{3}{2} \frac{\Delta v^2}{\Delta u^2} \right).
\end{equation}
The corresponding integrated probabilities for the three collisional outcomes are
\begin{align}
  \begin{split}
    p_{\rm f} &= \int^{+\infty}_{u_{\rm f}} P(\Delta v | \Delta u) \textnormal{d} \Delta v \\
    &= 1 + \sqrt{\frac{6}{\pi}} \exp\left(-\frac{3}{2} \frac{u_{\rm f}^2}{\Delta u^2} \right) \frac{u_{\rm f}}{\Delta u} - \text{erf} \left(\sqrt{\frac{3}{2}} \frac{u_{\rm f}}{\Delta u}\right),
  \end{split} \\
  \begin{split}
    p_{\rm c} &= \int^{u_{\rm b}}_{0} P(\Delta v | \Delta u) \textnormal{d} \Delta v \\
    &= - \sqrt{\frac{6}{\pi}} \exp\left(-\frac{3}{2} \frac{u_{\rm b}^2}{\Delta u^2} \right) \frac{u_{\rm b}}{\Delta u} + \text{erf} \left(\sqrt{\frac{3}{2}} \frac{u_{\rm b}}{\Delta u} \right),
  \end{split} \\
  p_{\rm b} &= 1 - p_{\rm c} - p_{\rm f},
\end{align}
where $u_{\rm f}$ and $u_{\rm b}$ are the threshold velocities for fragmentation and bouncing, $p_{\rm b}$ is the probability for bouncing.  Throughout this paper, we adopt $u_{\rm b} = 5$ cm s$^{-1}$.

Furthermore, our models take into account the effects of cratering and modulated mass transfer in addition to destructive fragmentation (see the three sub-categories of fragmentation outcomes in Figure \ref{fig:scenarios}).  When the mass ratio between the target particle and the projectile 
particle ($q\equiv m_{\rm tg} / m_{\rm pj}$) in a fragmentation event exceeds $10$, we assume that cratering takes place, where the projectile excavates a certain amount of mass from the target, leaving the majority of the target intact.  The removed mass from the target is assumed to be the same as $m_{\rm pj}$.  Thus, a total amount of $2m_{\rm pj}$ fragments are distributed to particles with masses smaller than $m_{\rm pj}$ according to Equation \ref{eq:S_power_law}.  When $q > 50$, we instead assume that mass transfer happens in fragmentation, where $0.1 m_{\rm pj}$ is integrated into the target, with the rest $0.9 m_{\rm pj}$ fragments to smaller particles. To make the transition from cratering to mass transfer smooth, we use a simple spline function
\begin{equation}
  f_{\rm frag}(q) = \left\{\begin{aligned}
    &2\ &10 < q \leqslant 15 \\
    &1.45 + 0.55 \cos{\left(\frac{q - 15}{50 - 15}\pi\right)} \ &15<q\leqslant50 \\
    &0.9\ &q>50
  \end{aligned}\right. ,
\end{equation}
to determine the amount of mass, in units of $m_{\rm pj}$, that becomes fragments and then the corresponding mass that is removed or added to the target particle.

\citet{Windmark2012} adopted a more elaborate model to determine collisional outcomes, where the threshold velocities for bouncing and fragmentation vary with particle size.  In this work, we stick to our simple and straightforward treatment such that fragmentation velocity threshold can be easily parameterized in our survey simulations and defer the more complicated treatment for future work.

To limit the artificial growth of mass bins with unrealistic low number densities and avoid ``lucky'' breakthrough to larger particles due to velocity distribution and mass transfer, we follow \citet{Drazkowska2014} and include a modulation function ($f_{\rm mod}$) in the coagulation kernel
\begin{equation}
  \begin{aligned}\label{eq:Kmod}
    K_{\rm mod}(m_1, m_2) &= K(m_1, m_2) f_{\rm mod}(m_1, m_2) \\
    &= K(m_1, m_2) \exp\left(-\frac{1}{N_1} - \frac{1}{N_2}  \right),
  \end{aligned}
\end{equation}
where $N_1$ and $N_2$ are numbers of particles in a $0.1 R_{\rm accu}$ wide annulus.  Since artificial growth may also take place during fragmentation events with large mass ratios (e.g., via mass transfer; see also Appendix \ref{appsubsec:more_tests}), it is \textit{necessary} to include the same modulation function in the fragmentation kernel as well
\begin{align}\label{eq:Lmod}
    L_{\rm mod}(m_1, m_2) &= L(m_1, m_2) f_{\rm mod}(m_1, m_2).
\end{align}
Below we consider modulation on both kernels unless otherwise stated. 

%%%%%%%%%%%%%%%%%%%%%%%%%%%%%%%%%%%%%%%%%%%%%%%%%%
\subsubsection{Dust Feedback on Turbulent Diffusion}
\label{subsubsec:feedback}

When solids begin to pile-up and dominate the local mass, the back reaction of particles to the gas becomes non-negligible \citep{Hyodo2019, Hyodo2021, Ida2021}.  To account for such feedback effect in dust diffusion in gas, we parameterize the reduction of diffusivity by replacing $\alpha$ with
\begin{equation}\label{eq:alpha_FB}
    \alpha_{\rm FB} \equiv \frac{\alpha}{\left(1 + \epsilon_{\rm mid}\right)^{\mathcal{K}}}
\end{equation}
in dust property calculations, where
\begin{equation}
  \epsilon_{\rm mid} = \frac{1}{\rho_{\rm rad}} \int_0^\infty \frac{\sigma(a)}{\sqrt{2\pi}H_{\rm d}} \textnormal{d}\log a
\end{equation}
is the total dust-to-gas density ratio in the disc midplane. 
%\yxc{perhaps we can specify how we calculate this?} 
The value of $\mathcal{K}$ is debatable and is in active study \citep{Ida2021}.  In this work, we adopt $\mathcal{K}=1$ for simplicity.

The feedback effect is only non-negligible when $\epsilon_{\rm mid} \gg 1$, where the dust scale heights and the relative turbulent velocities between particles are reduced by a factor of $\sim \sqrt{1+\epsilon_{\rm mid}}$.  Given the uncertainties of $\mathcal{K}$ and the lack of knowledge for heavy dust-loading scenarios, the maximum feedback effect is numerically limited by
\begin{equation}\label{eq:alpha_FB_100}
    \alpha_{\rm FB} = \frac{\alpha}{\max[\left(1 + \epsilon_{\rm mid}\right), 100]}.
\end{equation}
We again defer a more comprehensive understanding of $\alpha_{\rm FB}$ to future work.

%%%%%%%%%%%%%%%%%%%%%%%%%%%%%%%%%%%%%%%%%%%%%%%%%%
\subsection{Dust Evolution due to Accretion}
\label{subsec:dust_evo}

To model the dust evolution at $R_{\rm accu}$ with dynamic mass exchange with the environment, we take into account the supplementary solids drifted in from external discs as well as small particles that are carried away by accretion funnels.

The dust supply is embedded in the disc accretion flow.  The gas surface density at $R_{\rm accu}$ is assumed to be a constant and in an equilibrium state with an accretion rate of $\dot{M}$ (i.e., disc accretion is balanced by accretion onto the star).  We then assume that $\dot{\Sigma}_{\rm g} = \dot{M} / (2 \pi R_{\rm accu} H)$ and the dust-to-gas ratio of the accreted materials from external disc is $Z_{\rm supp}$ such that $\dot{\Sigma}_{\rm d,in} = Z_{\rm supp} \dot{\Sigma}_{\rm g}$, where solids follow the MRN distribution and the maximum particle size is $a_{\rm supp, max}$. 

The dust distribution only lose particles smaller than $r_{\rm d,critD}$ (see Equation \ref{eq:r_d_critD} and Figure \ref{fig:r_crit_map}) to the accretion funnels.  The loss rate is assumed to be
\begin{equation}\label{eq:sigma_dot}
    \dot{\sigma}(a) = {\sigma(a)} f_{\rm H}(a) \frac {\dot{\Sigma}_{\rm g} }{\Sigma_{\rm g}},
\end{equation}
where $f_{\rm H}(a)$ denotes the fraction of dust mass above the funnel launching point (i.e., one gas scale height; see Section \ref{subsec:funnel_flows}) and the surface density loss rate is $\dot{\Sigma}_{\rm d,out} = \int \dot{\sigma}(a)\ d\log a$.

If the solid distribution reaches a steady state without mass accumulation, the final dust-to-gas surface density ratio $Z_{\rm final}$ may be estimated by assuming that all the supplied dust is pulverized to the smallest dust species and carried away by gas accretion
\begin{equation}
    \left. \begin{aligned}
        \frac{\dot{\sigma}}{\sigma} = \frac{\dot{\Sigma}_{\rm d,out}}{\Sigma_{\rm d}} &= \frac{\dot{\Sigma}_{\rm g}}{\Sigma_{\rm g}} f_H \\
        Z_{\rm supp} \dot{\Sigma}_{\rm g} &= \dot{\Sigma}_{\rm d,out}
    \end{aligned} \right\}
    \Rightarrow Z_{\rm final} = \frac{Z_{\rm supp}}{f_H}, \label{eq:pulverized_Z_final}
\end{equation}
where $f_H = 1 - \text{erf}(1/\sqrt{2}) \sim 0.3$.  Thus, 
\begin{equation}\label{eq:Z_final}
    Z_{\rm final} \simeq 0.033 \left(\frac{Z_{\rm supp}}{0.01}\right).
\end{equation}

%%%%%%%%%%%%%%%%%%%%%%%%%%%%%%%%%%%%%%%%%%%%%%%%%%
\begin{table}
  \caption{Simulation Parameters \label{tab:paras}}
  \begin{tabular}{c|c|c|c|c|c}
  \hline
  Setup &
  $\alpha$ &
  $H$ &
  $T_{\rm rad}$ &
  $\Sigma_{\rm g}$ & 
  $r_{\mathrm{d,critD}}$ \\
  &
  &
  [au] &
  [K] &
  [g cm$^{-2}$] &
  [cm] \\
  \hline\hline
  A & 1.00e-3 & 0.00313 & 1880.60 & 1451.70 & 1.47e-02 \\
  B & 1.78e-3 & 0.00296 & 1676.09 &  915.96 & 1.56e-02 \\
  C & 3.16e-3 & 0.00279 & 1493.82 &  577.93 & 1.65e-02 \\
  D & 5.62e-3 & 0.00264 & 1331.37 &  364.65 & 1.75e-02 \\
  E & 1.00e-2 & 0.00249 & 1186.58 &  230.08 & 1.86e-02 \\
  \hline\hline
  \multicolumn{6}{c}{} \\[-0.5em]
  \multicolumn{6}{c}{All available choices$^{*}$ of $u_{\rm f}$, $a_{\rm supp,max}$, and $Z_{\rm supp}$} \\[0.25em]
  \hline
  \multicolumn{2}{c|}{$u_{\rm f}$ [cm s$^{-1}$]} &
  \multicolumn{4}{c}{100, 178, 316, 562, 1000} \\
  \multicolumn{2}{c|}{$a_{\rm supp,max}$ [cm]} &
  \multicolumn{4}{c}{30, 100, 300, 1000} \\
  \multicolumn{2}{c|}{$Z_{\rm supp}$} &
  \multicolumn{4}{c}{0.01, 0.05} \\
  \hline
  \end{tabular} \\
  {\large N}OTE --- For all runs, $R_{\star}=1.8 R_\odot$, $B_{\star}=1.0$ kG, $P_{\star}$ = 8 days, $f_{\rm out} = 1.25$, and thus $\dot{M} = 3.06$e$-9 M_\odot$ yr$^{-1}$ , $R_{\rm accu} = 0.098$ au.  Our simulations explore all possible combinations ($200$ in total) of key parameters ($\alpha$, $u_{\rm f}$, $a_{\rm supp,max}$, $Z_{\rm supp}$). 
\end{table}

\begin{table*}
  \caption{Fiducial Models \label{tab:fiducial}}
  \begin{tabular}{l|c|c|c|c|c|c}
  \hline
  Model Name$^{*}$ &
  $\alpha$ &
  $u_{\rm f}$ &
  $a_{\rm supp, max}$ &
  $Z_{\rm supp}$ &
  Mass Transfer &
  Feedback Effect \\
   &
   &
  [cm s$^{-1}$] &
  [cm] &
   &
   &
  \\
  \hline\hline
  %\rule{0pt}{1cm} % used if need to preserve space in this row
  A1\_MT     & \multirow{6}{*}{$1.0$e$-3$} & \multirow{6}{*}{$1000$} & \multirow{6}{*}{$100$} & \multirow{3}{*}{$0.01$} 
                             & \checkmark &            \\
  %\cline{1-1}\cline{6-7}
  A1\_FB     &   &   &   &   &            & \checkmark \\
  %\cline{1-1}\cline{6-7}
  A1\_MT\_FB &   &   &   &   & \checkmark & \checkmark \\
  \cline{1-1}\cline{5-7}
  A5\_MT     &   &   &   & \multirow{3}{*}{$0.05$} 
                             & \checkmark &            \\
  %\cline{1-1}\cline{6-7}
  A5\_FB     &   &   &   &   &            & \checkmark \\
  %\cline{1-1}\cline{6-7}
  A5\_MT\_FB &   &   &   &   & \checkmark & \checkmark \\
  \hline
  B1\_MT     &  \multirow{6}{*}{$1.78$e$-3$} & \multirow{6}{*}{$1000$} & \multirow{6}{*}{$1000$} & \multirow{3}{*}{$0.01$}
                             & \checkmark &            \\
  %\cline{1-1}\cline{6-7}
  B1\_FB     &   &   &   &   &            & \checkmark \\
  %\cline{1-1}\cline{6-7}
  B1\_MT\_FB &   &   &   &   & \checkmark & \checkmark \\
  \cline{1-1}\cline{5-7}
  B5\_MT     &   &   &   & \multirow{3}{*}{$0.05$}
                             & \checkmark &            \\
  %\cline{1-1}\cline{6-7}
  B5\_FB     &   &   &   &   &            & \checkmark \\
  %\cline{1-1}\cline{6-7}
  B5\_MT\_FB &   &   &   &   & \checkmark & \checkmark \\
  \hline
  \end{tabular} \\
  \begin{flushleft}
    $^{*}$ ``MT'' indicates mass transfer (with modulated kernels) is enabled. ``FB'' indicates feedback effect is enabled.  ``MT\_FB'' indicates both mechanisms are enabled. \\
  \end{flushleft}
\end{table*}

%%%%%%%%%%%%%%%%%%%%%%%%%%%%%%%%%%%%%%%%%%%%%%%%%%
\subsection{Numerical Setup}
\label{subsec:setup}

Table \ref{tab:paras} summarizes the physical and numerical parameters for our dust evolution simulations.  In all cases, we assume $R_{\star} = 1.8 R_\odot$, $B_{\star} = 1.0$kG, and $P_{\star} = 8$ days, which leads to $\dot{M} = 3.06\times10^{-9} M_\odot$ yr$^{-1}$ and $R_{\rm accu} = 0.098$ au.  The dust evolution is then controlled by four parameters of our interest: $\alpha$, $u_{\rm f}$, $a_{\rm supp,max}$, and $Z_{\rm supp}$.

The first parameter $\alpha$ is the only remaining free parameter for constructing a radiative disc profile (i.e., determining $H$, $T_{\rm rad}$, and $\Sigma_{\rm g}$).  We vary $\alpha$ from $1$e-$3$ to $1$e-$2$ to roughly cover the temperature regime of interest, i.e., between $1000$ K and $2000$ K.  The choice of $\alpha$ also affects dust scale height and relative velocities (via $\alpha_{\rm FB}$; see Equation \ref{eq:alpha_FB_100}).  Since $T_{\rm rad}$ and $c_{\rm s}$ decreases with $\alpha$ (see Equation \ref{eq:T_rad}), the final dust relative velocities depend on the synergy of $T_{\rm rad}$ and $\alpha$.  

The second parameter $u_{\rm f}$ determines how easily particles fragment upon collisions and affects how efficient grains are able to grow.  The nominal fragmentation velocities for silicate grains identified in previous studies are of the order of a few m s$^{-1}$ \citep[e.g., ][]{Blum2008}.  Thus, we vary $u_{\rm f}$ from 100 to 1000 cm s$^{-1}$ to compensate the uncertainty.

The two remaining parameters $a_{\rm supp,max}$ and $Z_{\rm supp}$ govern the size range and the amount of dust supply (relative to gas) from external disc.  In this work, we consider a wide range of $Z_{\rm supp}$ ($0.01$ and $0.05$) and $a_{\rm supp, max}$ ($30$, $100$, $300$, and $1000$ cm) due to ambiguities in the realistic disc conditions relevant for planetesimal formation.  First, mass ratios of gas and solids in discs are poorly determined quantities in observations since accessible mass-sensitive tracers/species are limited and mass estimation is associated with various assumptions (e.g., abundance, optical depth, temperature) \citep[][etc.]{Andrews2015, McClure2016, Bergin2017, Zhu2019, Andrews2020}.  By design, the derived masses for both gas and solids are likely lower bounds.  Consequently, the estimates of bulk dust-to-gas ratios in protoplanetary discs are intrinsically uncertain and spread out over a wide range \citep[e.g.,][]{Ansdell2016, Ansdell2018, Miotello2017, Cieza2019}.

Even less is known of the specific dust size distribution of in-drifting particles.  Since the dust size corresponding to St=1 may decrease with disc radius when $\Sigma_{\rm g}$ becomes large enough for gas drag to enter the Stokes regime, making it easier for large particles to break through the drift barrier.  For example, Figure 11 of \citet{Birnstiel2010} shows that very large particles ($\gtrsim10^4$ cm) are likely produced at sub-au if the Stokes regime is included.  Therefore, we take into account that the supplied materials may include very large particles (i.e., $a_{\rm supp, max}$ up to $1000$ cm).  We discuss how our results depend on the choice of the maximum supplied grain size in Section \ref{subsec:survey_results}.

Our simulations traverse all the $200$ combinations of the four key parameters.  In each simulation, we model the dust size distribution from $10^{-4}$ to $10^{4}$ cm with 202 mass bins, which corresponds to $8.4$ grid points per mass decade, a typical resolution used in dust coagulation models.  Solids are initially monodisperse at $10^{-4}$ cm (i.e., all are $\mu$m-sized dust grains) and the dust-to-gas surface density ratio $Z \equiv \Sigma_{\rm d}/\Sigma_{\rm g}$ is initialized at $0.01$.  

At each end of the mass grid, there is one ghost mass bin representing the boundary conditions.  Since the dust distribution dynamically exchanges masses with the disc accretion flow, we adopt unilaterally active ghost bins in this work.  The ghost mass bin at the small mass end continues losing solids due to accretion funnels.  To mimic dust replenishment processes (e.g., condensation) that counter such loss, we allow the small-end ghost mass bin to coagulate with all bins.  Furthermore, the ghost mass bin at the large mass end is allowed to collide with all bins and experience fragmentation such that breakthrough particles can continue to retain dust mass via mass transfer.

Along the dust evolution, we continuously check for the conservation of the total dust surface density and enforce the precision of each timestep to be better than one part in a million \citep[i.e., $1$e$-6$, ][]{Garaud2013}.  In addition, we renormalize the total surface density in the entire domain in each timestep, after solving the Smoluchowski equation and before accounting for the mass gain/loss due to accretion, to minimize the influence of numerical errors on the evolution of $\Sigma_{\rm d}$.

Our goal is to understand the required conditions under which solids are able to accumulate efficiently and either become unstable to self-gravity, or to grow beyond the fragmentation barrier and produce planetesimals.  Motivated by this goal, we only evolve all the models for $10^5$ years and consider $Z \gtrsim 1$ as the criterion for significant solid accumulation.

%%%%%%%%%%%%%%%%%%%%%%%%%%%%%%%%%%%%%%%%%%%%%%%%%%
\begin{figure*}%[!ht]
  %\centering
  %\makebox[\textwidth][c]{}
  \includegraphics[width=\linewidth]{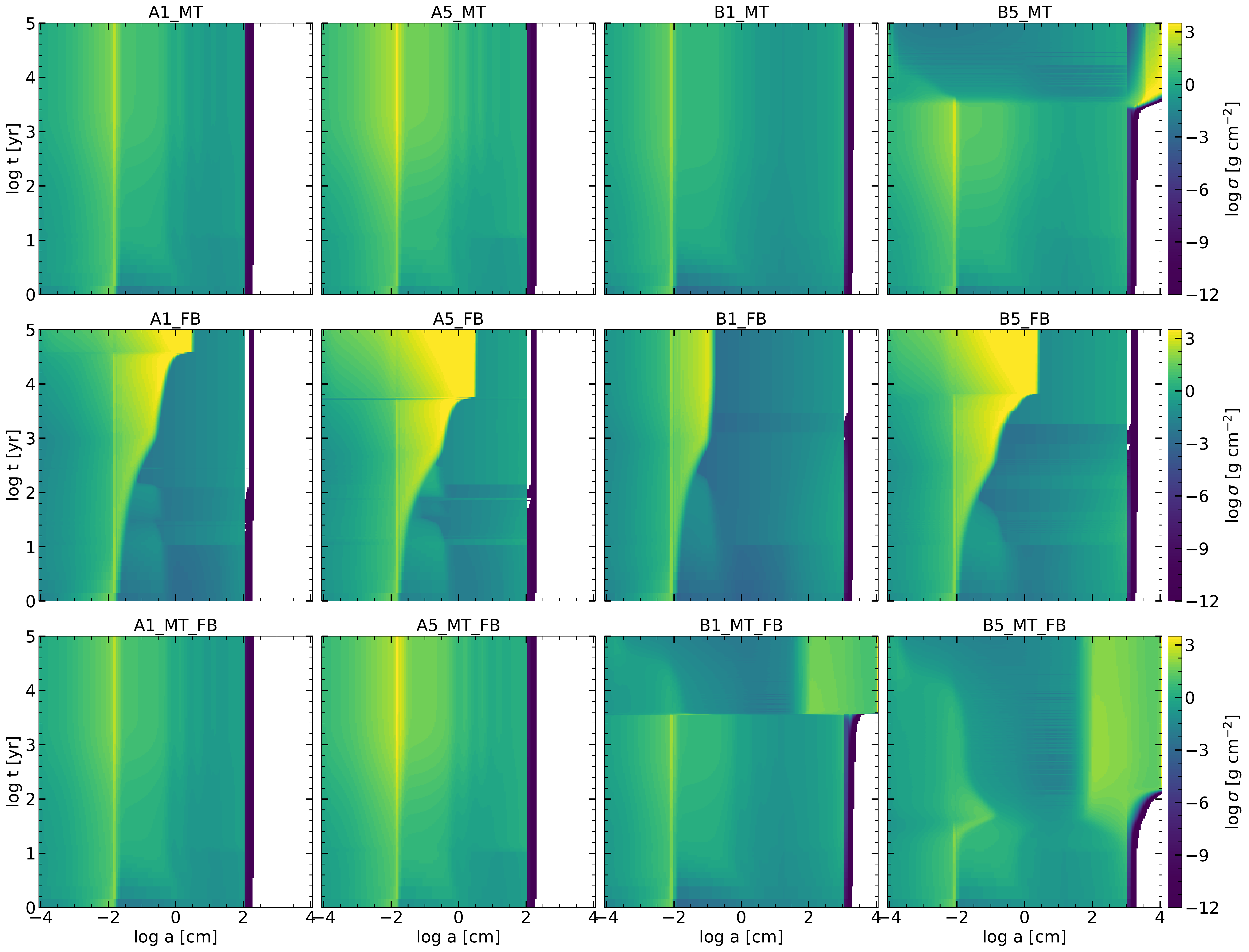}
  \caption{The evolution of vertically integrated dust surface density distribution in the particle size-time plane for all of our fiducial local dust evolution models (see Table \ref{tab:fiducial} and Section \ref{subsec:fiducial_cases}).  These cases are representative in that all possible outcomes of dust evolution are covered, including quasi-steady dust distribution without significant solid accumulation, heavy dust-loading in the mass reservoir of cm-sized dust via feedback effect, and runaway growth through mass transfer. 
  \label{fig:fiducial}}
\end{figure*}

\begin{figure*}%[!htb]
  \centering
    \includegraphics[width=0.9\linewidth]{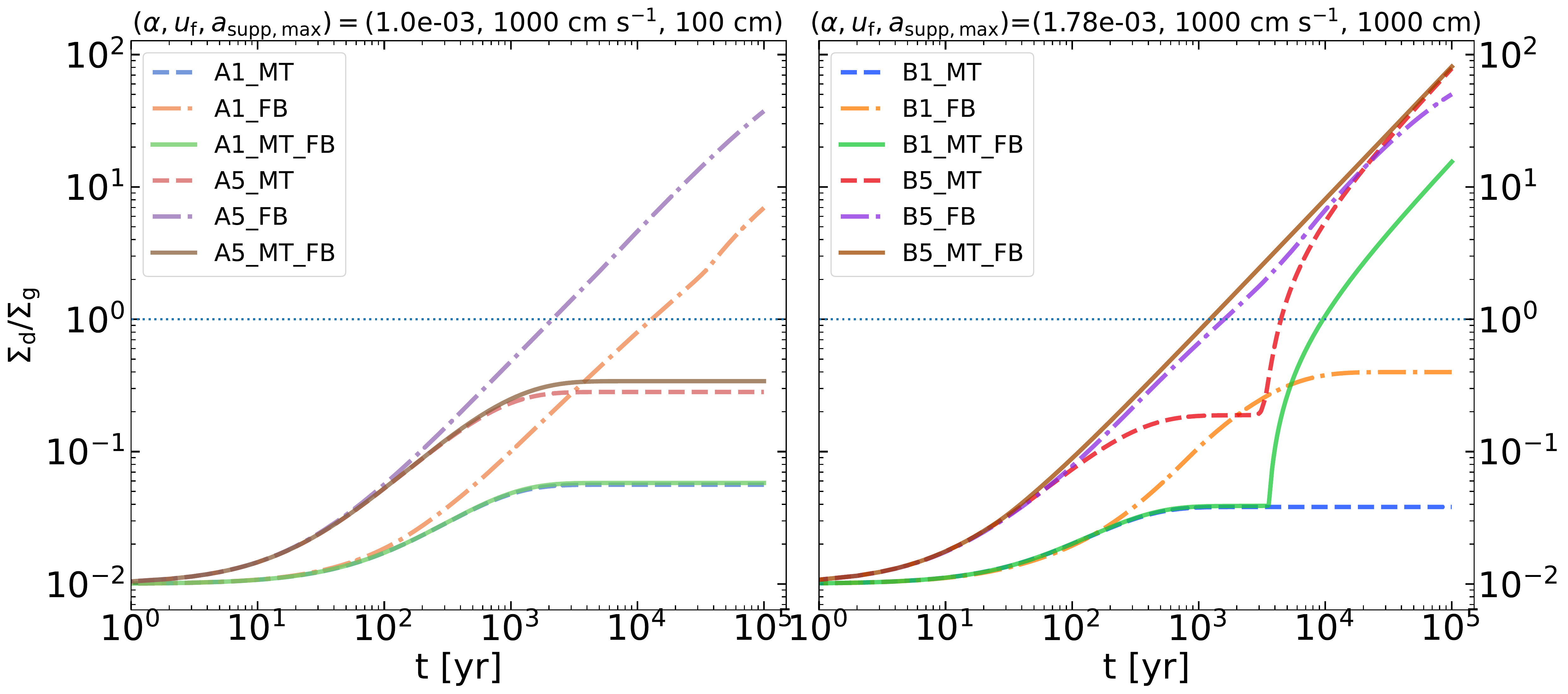}
  \caption{The evolution of dust-to-gas surface density ratio for all of our fiducial local dust evolution models (see Table \ref{tab:fiducial} and Section \ref{subsec:fiducial_cases}).  Note that $\Sigma_{\rm g}$ decreases with $\alpha$ such that the value in \texttt{A}-series is different than that in \texttt{B}-series.
  \label{fig:Z_fiducial}}
\end{figure*}

\begin{figure}%[htb]
  \centering
    \includegraphics[width=0.9\linewidth]{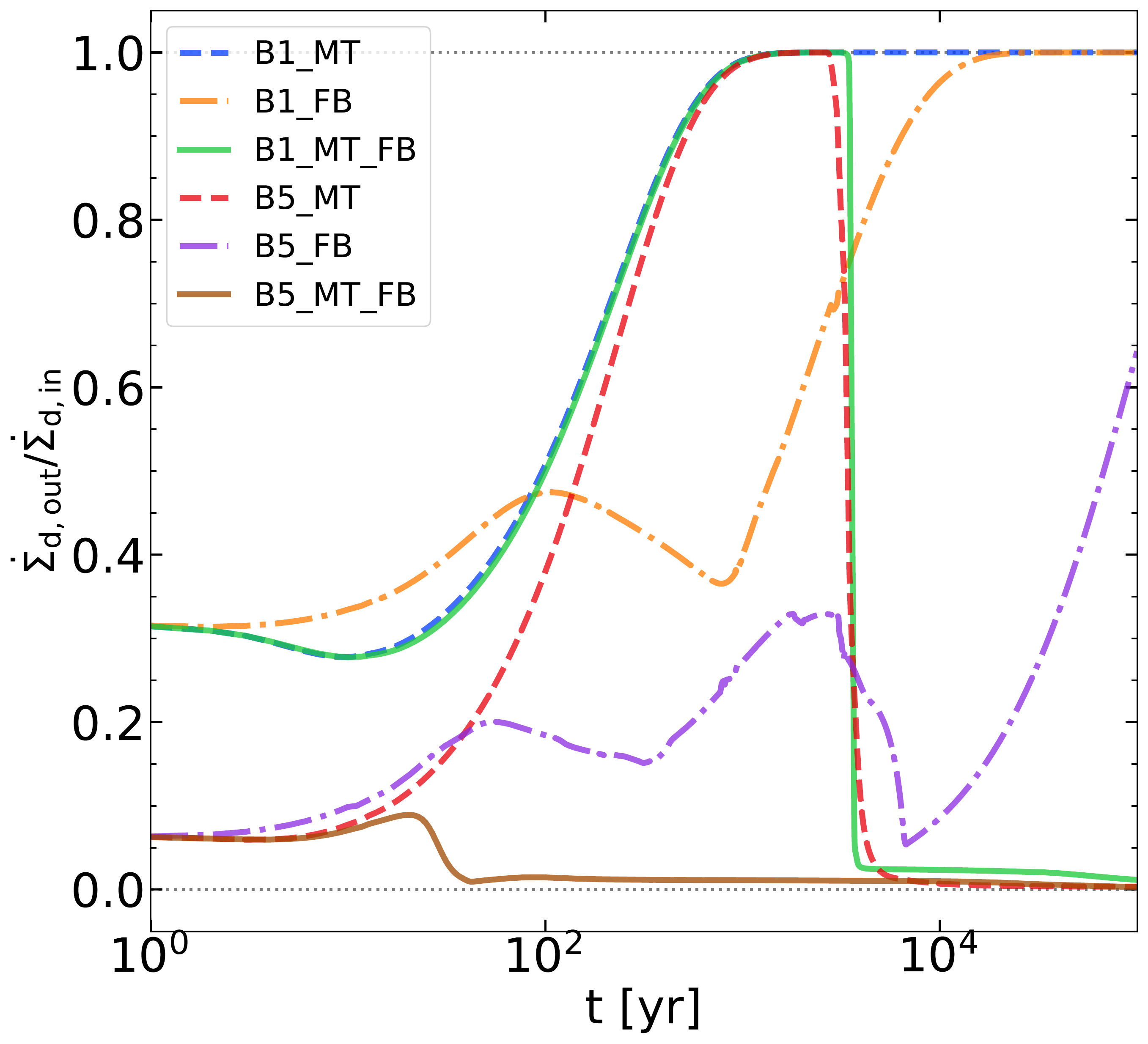}
  \caption{Dust removal efficiency for our fiducial models in \texttt{B}-series.  This efficiency shows the ratio between the dust loss rate due to accretion funnels ($\dot{\Sigma}_{\rm d,out}$) and the dust supply rate from disc accretion ($\dot{\Sigma}_{\rm d,in}$).  Line styles and colors have the same meaning as Figure \ref{fig:Z_fiducial}.  The dust removal efficiency varies in different dust evolution outcomes, where $1 - \dot{\Sigma}_{\rm d,out}/\dot{\Sigma}_{\rm d,out} = 1$ indicates the efficiency of dust retention from disc accretion.
  \label{fig:fiducial_dust_removal}}
\end{figure}

\begin{figure*}%[htb]
  \centering
    \includegraphics[width=0.9\linewidth]{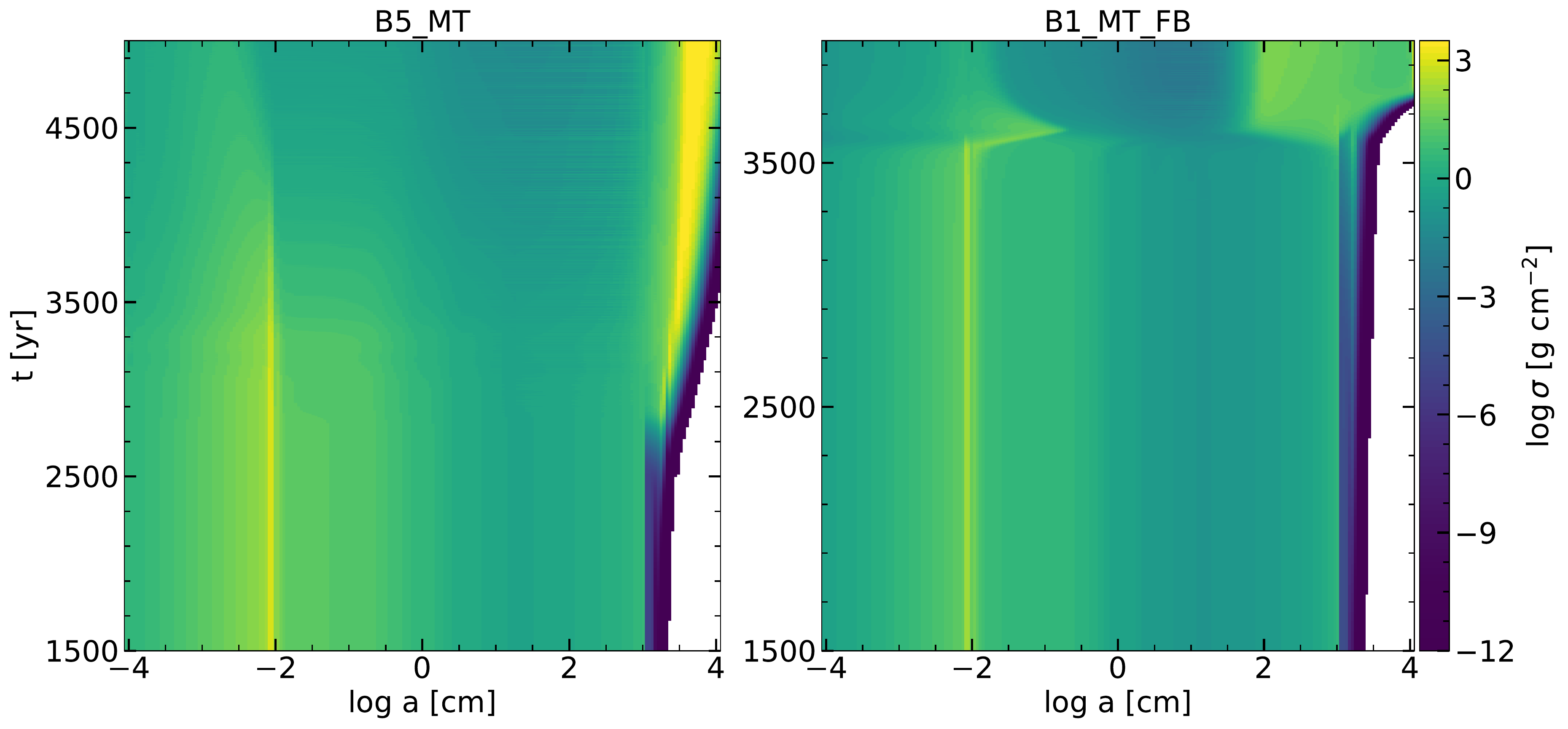}
  \caption{Similar to Figure \ref{fig:fiducial} but focusing on the runaway growth via mass transfer solely (\textit{left}) and via both mass transfer and feedback effects (\textit{right}).  The feedback effect assisted runaway growth is much faster regarding the time needed to produce particles towards $10^4$ cm (see Section \ref{subsubsec:coopetition}).
  \label{fig:fiducial_zoom_in}}
\end{figure*}

%%%%%%%%%%%%%%%%%%%%%%%%%%%%%%%%%%%%%%%%%%%%%%%%%%%%%%%%%%%%%%%%%%%%%%%%%%%%%%%%
\section{Results}
\label{sec:results}

We conduct a suite of dynamic local dust evolution simulations at $R_{\rm accu}$ with parameters listed in Table \ref{tab:paras} to study dust growth near the disc inner boundary.  In this section,  we first present a detailed analyses on a few fiducial cases in Section \ref{subsec:fiducial_cases} and identify crucial physical processes that contribute to solid accumulation.  Section \ref{subsec:survey_results} then surveys the disc conditions needed to produce efficient dust retention.

%%%%%%%%%%%%%%%%%%%%%%%%%%%%%%%%%%%%%%%%%%%%%%%%%%
\subsection{Dust Evolution Scenarios}
\label{subsec:fiducial_cases}

We are particularly interested in the two mechanisms that may affect the coagulation-fragmentation equilibrium and lead to dust growth and solid accumulation, namely mass transfer and feedback effects.  To better understand and disentangle the roles of these two mechanisms in dust evolution, we perform controlled experiments on four fiducial models listed in Table \ref{tab:fiducial} (\texttt{A}-series and \texttt{B}-series, respectively), with the two mechanisms of interest enabled individually and combined (three scenarios in total for each model).  These four models are considered representative because their results cover all types of outcome scenarios.

Figures \ref{fig:fiducial} and \ref{fig:Z_fiducial} show that the evolution of all three scenarios are very similar in the first one hundred years, where the dust distributions quickly populate a wide size spectrum (from $10^{-4}$ cm to $a_{\rm supp,max}$) due to the fragmentation cascade of the supplied dust and the coagulation growth of small dust grains.  The peak sizes of these distributions are around $10^{-2}$ cm, which is roughly the fragmentation barrier of the coagulation growth (see also $a_{\rm P}$ in Figure \ref{fig:BM_steady_state}).  In the following sections, we focus on each individual scenario to depict and distinguish their dust evolution afterward.

%%%%%%%%%%%%%%%%%%%%%%%%%%%%%%%%%%%%%%%%%%%%%%%%%%
\subsubsection{Mass Transfer Only}
\label{subsubsec:mt_only}

The mechanism of mass transfer is of great interest because it is a physically motivated grain growth process where massive particles are able to embrace bombardments of small grains, and acquire more mass even in the size regime beyond the fragmentation threshold.  Our fiducial models \texttt{A1\_MT}, \texttt{A5\_MT}, \texttt{B1\_MT}, and \texttt{B5\_MT} (hereafter \texttt{MT} models) focus on mass transfer and ignore feedback effect.  They are thus similar to traditional dust population models in previous studies.  

The first three \texttt{MT} models reach quasi-steady states in a few thousand years without significant dust growth or accumulation.  More specifically, the peak size of the dust distribution remains sub-mm (see the yellow streaks in Figure \ref{fig:fiducial}) and the dust-to-gas surface density ratio $Z$ remains below unity.  On the contrary, Model \texttt{B5\_MT} breaks through the fragmentation barrier and produces considerable runaway growth, both in particle size and total solid mass.

The dust evolution becomes quasi-steady in the first three models when the balance between the dust supply from the outer disc and the dust loss due to accretion funnels is achieved.  We define the ratio between the dust loss rate and the dust supply rate as the \textit{dust removal efficiency}, which is unity in the aforementioned quasi-steady state (i.e., $\dot{\Sigma}_{\rm d,out}/\dot{\Sigma}_{\rm d,in} = 1$; see Figure \ref{fig:fiducial_dust_removal}).  The final dust-to-gas surface density ratio $Z_{\rm final}$ in Model \texttt{B1\_MT} are in good agreement with our prediction: $\sim 3.3$ per cent for $Z_{\rm supp}=0.01$ (see Equation \ref{eq:Z_final}).  However, in Models \texttt{A1\_MT} and \texttt{A5\_MT} with relatively weaker turbulent diffusion, the values of $Z_{\rm final}$ are slightly higher than our simple prediction because more solids are needed to activate severer fragmentation to counteract coagulation and reach an equilibrium.

Solid growth to particles larger than $a_{\rm supp,max}$ takes place in all \texttt{MT} models (see the first row of Figure \ref{fig:fiducial}) due to the inclusion of velocity probability distribution in collision treatments (see Section \ref{subsubsec:collisional_outcomes}), where coagulation is still possible even with $\Delta u > u_{\rm f}$.  However, the amount of such particles is initially tiny such that mass transfer beyond $a_{\rm supp,max}$ is heavily modulated and is equivalently deactivated.  The primitive accumulation of these particles is thus gradual and solely depends on the efficiency of direct coagulation.

Among the \texttt{MT} models, only Model \texttt{B5\_MT} piles up enough large particles that alleviates the kernel modulation due to the relative large $a_{\rm supp,max}$ and $Z_{\rm supp}$.  Mass transfer is immediately activated for these large particles, which sweep up small dust grains and become larger and larger.  Consequently, the peak size of the particle distribution transfers from $\sim 10^{-2}$ cm to $\gtrsim 10^{3.5}$ cm (see Figure \ref{fig:fiducial_zoom_in}) and grain growth enters the size regime where self relative velocities drop below $u_{\rm f}$ (see Figure \ref{fig:dv_tot}), making direct coagulation viable and leading to runaway growth and hence breakthrough.  Since most of the solid mass now lies in the very large particles ($>10^{3.5}$ cm), mass transfer dominates the dust evolution and transports almost all the supplied dust mass into very large particles, leading to $\dot{\Sigma}_{\rm d,out}/\dot{\Sigma}_{\rm d,in} \simeq 0$ (see Figure \ref{fig:fiducial_dust_removal}).

The footprint of mass transfer is also seen in other \texttt{MT} models without breakthrough.  Taking the later quasi-steady state of Model \texttt{A5\_MT} as an example, mass transfer continuously transports mass from dust grains at the peak size ($\sim 10^{-2}$ cm) of the size distribution to larger particles, resulting an evenly-spaced chain of $\sigma(a)$ enhancements with a step length of roughly $50$ times in mass ($\sim 0.5$ in logarithmic size; see the vertical strips in Figure \ref{fig:fiducial}).  Such a step length is numerically determined by the selected mass ratio for mass transfer to become fully effective (see also Appendix \ref{appsubsec:more_tests}).  The dependency of dust evolution on this mass ratio is beyond the scope of this work and remains a topic for future studies.  That said, the reason that further dust growth does not takes place within the simulation time in Model \texttt{A5\_MT} is largely due to the high speed impacts between particles near $a_{\rm supp,max}$, making the accumulation of solids beyond $a_{\rm supp,max}$ inefficient.
  
%%%%%%%%%%%%%%%%%%%%%%%%%%%%%%%%%%%%%%%%%%%%%%%%%%
\subsubsection{Feedback Effect Only}
\label{subsubsec:fb_only}

The feedback effect (see Section \ref{subsubsec:feedback}), to our knowledge, has not been considered in local dust coagulation models in previous works.  It is however well motivated and raises an interesting possibility that heavy dust-loading reduces dust diffusion and facilitates coagulation growth.  Our fiducial models \texttt{A1\_FB}, \texttt{A5\_FB}, \texttt{B1\_FB}, and \texttt{B5\_FB} (hereafter \texttt{FB} models) focus on feedback effect and consider cratering for any fragmentation events where the mass ratio of the colliding particles exceeds $10$ (i.e., ignore mass transfer).

These \texttt{FB} models accumulate dust with sizes around cm to varying degrees depending on disc conditions.  If the pile-up of sub-cm particles is efficient such that feedback effect continuously reduces turbulent diffusion and thus relative velocities between dust (i.e., narrowing the diagonal destructive gap in Figure \ref{fig:dv_tot}), coagulation growth beyond the peak size of the early distributions becomes viable.  Consequently, the amount of pulverized dust carried away by funnel flows is reduced and significant solid accumulation is feasible in the mass reservoir of cm-sized solids, which in turn enhance the feedback effect.

The optimal disc conditions for the feedback effect include relatively weaker turbulent diffusion (lower $\alpha$ and lower $T_{\rm rad}$), larger $u_{\rm f}$, and larger $Z_{\rm supp}$.  All models in this scenario except Model \texttt{B1\_FB} experience strong feedback effect, where the peak size of the dust distribution reaches $\sim$cm and the dust-to-gas ratio increases monotonically throughout the simulation (see Figure \ref{fig:Z_fiducial}).  In their late evolution, dust accumulates so much that the regulation on the maximum feedback effect (see Equation \ref{eq:alpha_FB_100}) is activated, where the growth of peak size ceases at a few cm.

Model \texttt{B1\_FB} exhibits modest feedback effect that results in some initial dust accumulation at the early stage, where the minimum dust removal efficiency is $\sim 40$ per cent (see Figure \ref{fig:fiducial_dust_removal}).  Toward the end of the simulation, this model approaches a quasi-steady state with a near-unity removal efficiency, indicating the mass reservoir is almost full.  However, Model \texttt{B5\_FB} with $Z_{\rm supp}=0.05$ accumulates much more solid mass at the end, suggesting that faster dust supplementation can dynamically support a larger capacity of the mass reservoir.

In these \texttt{FB} models, the supplied solids eventually fragment into smaller grains due to destructive high speed collisions and the lack of mass transfer.  Models with different $a_{\rm supp, max}$ thus have degenerate results since it is the efficient coagulation growth that retains solid mass in grains near the peak size (i.e., $\sim$ cm, so $\gg r_{\rm crit,D}$) of the dust distribution.  We discuss more on this degeneracy and justify it in Section \ref{subsec:survey_results}.

%%%%%%%%%%%%%%%%%%%%%%%%%%%%%%%%%%%%%%%%%%%%%%%%%%
\begin{figure*}%[!ht]
  \centering
    \makebox[\textwidth][c]{ % enables figure wider than textwidth
    \includegraphics[width=1.\linewidth]{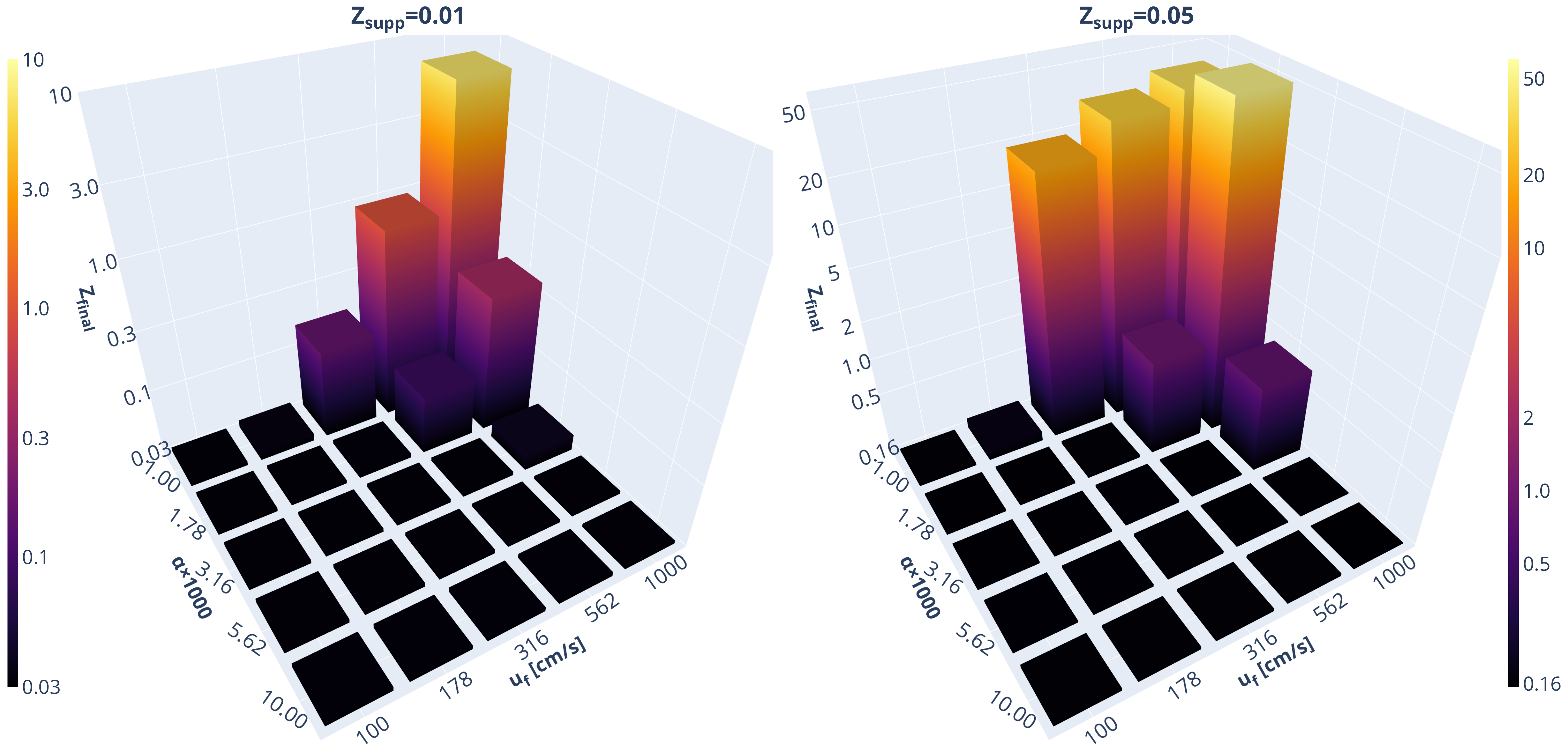}
    }
  \caption{Final dust-to-gas surface density ratios $Z_{\rm final}$ after $10^5$ years of local dust evolution for \texttt{FB} models as a function of $\alpha$, $u_{\rm f}$, and $Z_{\rm supp}$.  This figure neglects the dependency on $a_{\rm supp,max}$ since dust mass accumulates in the mass reservoir of cm-sized particles, resulting in degeneracy of $a_{\rm supp,max}$.  Solid accumulation is fast and robust under the optimal disc conditions, i.e., with low turbulent diffusion, a high fragmentation velocity threshold, and a large maximum supply particle size.  An interactive version of this plot is available at \href{https://rixinli.com/RubbleSurveyResults.html}{https://rixinli.com/RubbleSurveyResults.html}.
  \label{fig:Zfinal_capFB_noMT}}
\end{figure*}

\begin{figure*}%[!ht]
  \centering
    \makebox[\textwidth][c]{ % enables figure wider than textwidth
    \includegraphics[width=1.\linewidth]{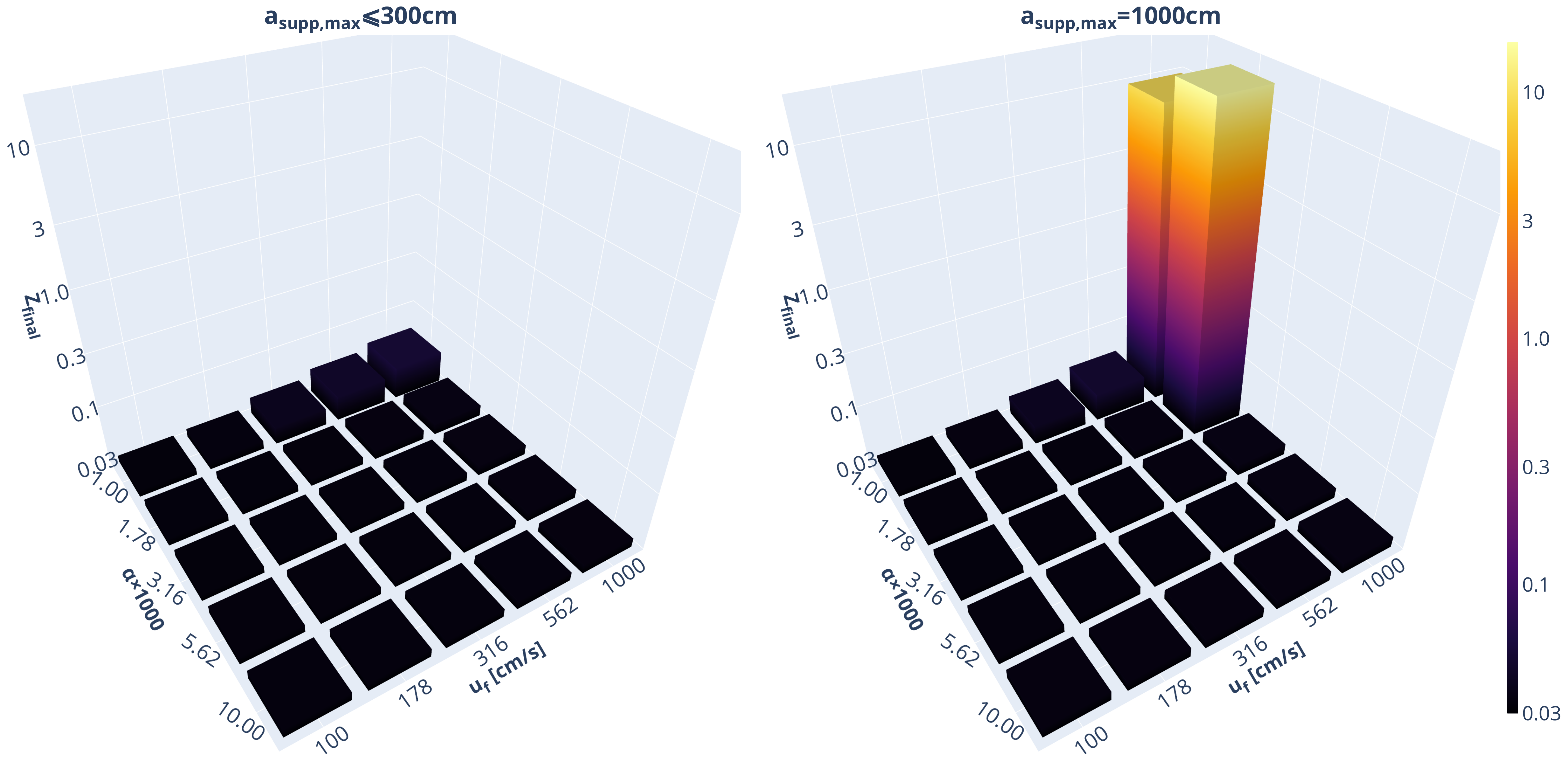}
    }
  \caption{Similar to Figure \ref{fig:Zfinal_capFB_noMT} but for \texttt{MT\_FB} models with $Z_{\rm supp}=0.01$.  Models with $a_{\rm supp,max} \leqslant 300$ cm (\textit{left}) do not accumulate considerable dust mass and produce almost identical results.  Models with $a_{\rm supp,max} = 1000$ cm (\textit{right}) may yield runaway growth and significant solid accumulation given ideal disc conditions.   An interactive version of this plot is available at \href{https://rixinli.com/RubbleSurveyResults.html}{https://rixinli.com/RubbleSurveyResults.html}
  \label{fig:Zfinal_capFB_MT}}
\end{figure*}

\begin{figure*}%[!ht]
  \centering
    \makebox[\textwidth][c]{ % enables figure wider than textwidth
    \includegraphics[width=1.\linewidth]{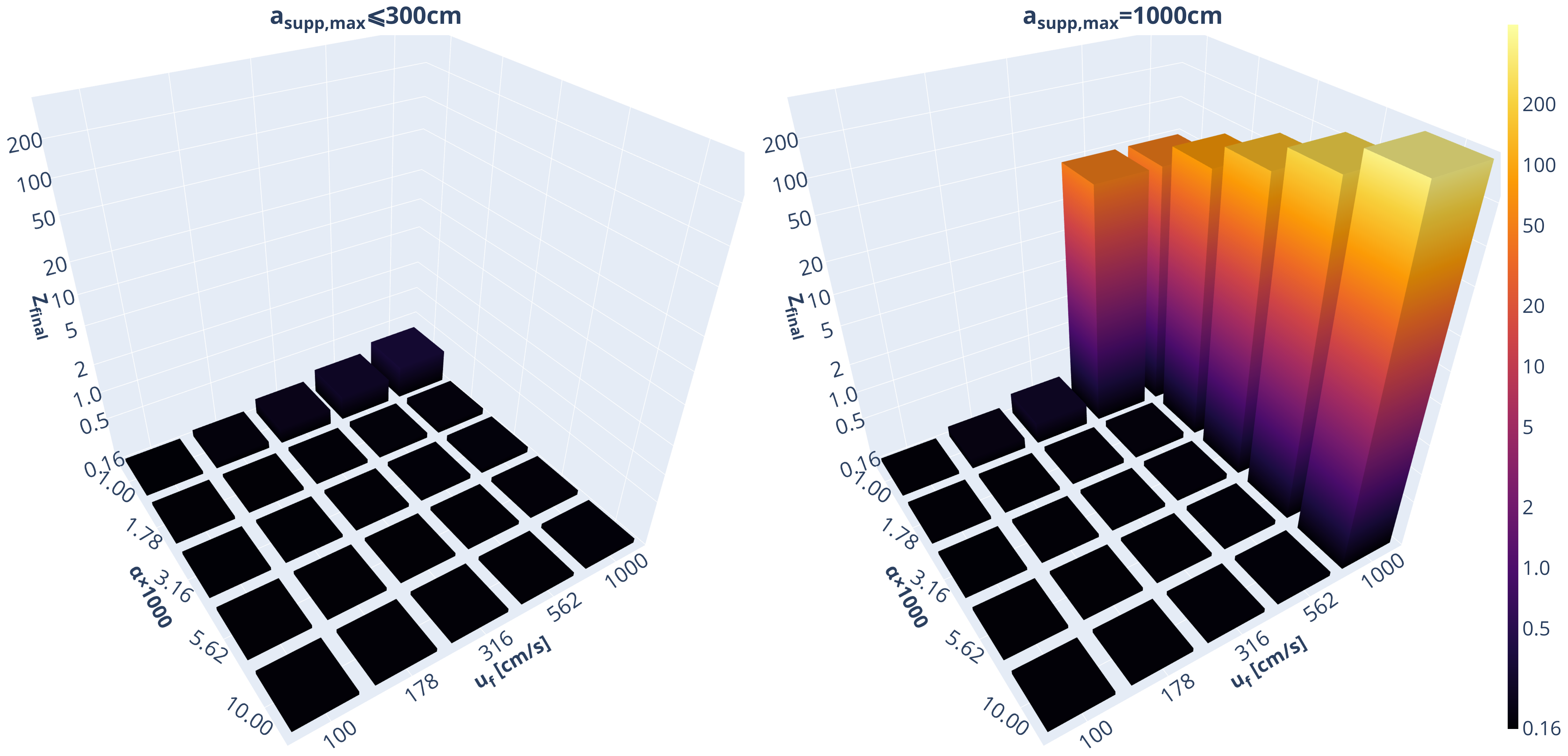}
    }
  \caption{Similar to Figure \ref{fig:Zfinal_capFB_MT} but for \texttt{MT\_FB} models with $Z_{\rm supp}=0.05$.  An interactive version of this plot is available at \href{https://rixinli.com/RubbleSurveyResults.html}{https://rixinli.com/RubbleSurveyResults.html}
  \label{fig:Zfinal_capFB_MT_Z5}}
\end{figure*}

%%%%%%%%%%%%%%%%%%%%%%%%%%%%%%%%%%%%%%%%%%%%%%%%%%
\subsubsection{``Co-opetition'' between Two Mechanisms}
\label{subsubsec:coopetition}

We are particularly interested in models with both mass transfer and feedback effects since these two mechanisms cooperate and compete with each other simultaneously (so-called ``co-opetition'').  Our fiducial models \texttt{A1\_MT\_FB}, \texttt{A5\_MT\_FB}, \texttt{B1\_MT\_FB}, and \texttt{B5\_MT\_FB} (hereafter \texttt{MT\_FB} models) take into account both mechanisms.  This paper is again to our knowledge the first work to investigate such models.

The behaviors of the first two \texttt{MT\_FB} models with $a_{\rm supp,max} = 100$ cm are similar to their \texttt{MT} model counterparts, where a quasi-steady state is reached in a few thousand years without significant dust growth or accumulation.  The peak size of the dust distribution remains sub-mm since mass transfer efficiently channels solid mass into larger particles (from cm to $a_{\rm supp,max}$).  Therefore, the feedback effect is largely damped compared to their \texttt{FB} model counterparts since the relative deficit of sub-mm-sized dust hinders the growth of the peak size in the particle distribution towards cm -- the size where solid accumulation is more efficient as shown in Section \ref{subsubsec:fb_only}.  Nevertheless, the fact that $Z_{\rm final}$ is slightly higher than that in their \texttt{MT} model counterparts still indicates the existence of feedback effect.  In this scenario, the two mechanisms compete with each other for solid mass and achieve a dynamic balance in the final quasi-steady state.

For dust evolution in models \texttt{B1\_MT\_FB} and \texttt{B5\_MT\_FB} with $a_{\rm supp,max} = 1000$ cm, the cooperation between feedback effects and mass transfer leads to breakthrough and runaway growth and eventually significant solid accumulation.  As explained in Section \ref{subsubsec:mt_only}, the key to runaway growth is the efficient grain growth into the size regime where self relative velocities drop below $u_{\rm f}$.  Instead of generating larger and larger particles, the feedback effect is able to reduce the relative velocities of all particles, making the required size regime for runaway growth smaller.  Moreover, the feedback effect also makes the accumulation of solids beyond $a_{\rm supp,max}$ easier and facilitate the activation of mass transfer.

In these two specific cases, the relative velocities between particles near $a_{\rm supp,max}$ are already marginally below $u_{\rm f} = 1000$ cm s$^{-1}$ (see Figure \ref{fig:dv_tot}).  With the assistance of feedback effect, runaway growth is much faster than that in \texttt{MT} models (see Figure \ref{fig:fiducial_zoom_in}).  The instant activation of mass transfer in turn strengthens the feedback effect, which establishes a positive feedback loop that depletes dust smaller than cm and produces particles towards $10^4$ cm almost immediately (see also Figures \ref{fig:Z_fiducial} and \ref{fig:fiducial_zoom_in}).  This growth takes place in a runaway manner because the larger the particles are, the smaller their relative velocities are, the lower their scale heights become and thus the stronger the feedback effect is.  After the runaway growth, Figure \ref{fig:fiducial_dust_removal} again shows that the mass transfer dominates the dust evolution and is able to capture and consume all the dust mass supplied from the outer disc.

%%%%%%%%%%%%%%%%%%%%%%%%%%%%%%%%%%%%%%%%%%%%%%%%%%
\subsection{Threshold for Significant Dust Accumulation}
\label{subsec:survey_results}

In this section, we summarize our survey results on all the combinations of key parameters, $\alpha$, $u_{\rm f}$, $a_{\rm supp,max}$, and $Z_{\rm supp}$ listed in Table \ref{tab:paras}.  Motivated by the findings that feedback effect  is vital for efficient dust accumulation and can assist the mass transfer mechanism to produce breakthrough, from now on we only focus on two scenarios, one with feedback effect only (i.e., \texttt{FB} models) and the other with both mechanisms (i.e., \texttt{MT\_FB} models).  

Figure \ref{fig:Zfinal_capFB_noMT} presents our survey results on the \texttt{FB} models and shows which runs retain substantial dust mass and which do not.  We neglect the dependency on $a_{\rm supp,max}$ and only present the results based on $a_{\rm supp,max} = 100$ cm since the results are almost identical for different $a_{\rm supp,max}$.  For example, $Z_{\rm final}$ in our fiducial models \texttt{B1\_FB} and \texttt{B5\_FB} with $a_{\rm supp,max} = 1000$ cm are $6.97$ and $50.16$, respectively.  The corresponding models presented in Figure \ref{fig:Zfinal_capFB_noMT} (with the same $\alpha$, $u_{\rm f}$, and $Z_{\rm supp}$) end with $Z_{\rm final} = 6.95$ and $50.03$, respectively, very close to those from our fiducial models.  As mentioned in Section \ref{subsubsec:fb_only}, the choice of $a_{\rm supp,max}$ has little influence on the dust evolution in this scenario because other disc conditions determine the coagulation efficiency, which then determine whether the feedback loop between dust mass retention in $\sim$ cm-sized grains and strong feedback effect can be activated or not.

To validate this interpretation, we perform additional experiments on \texttt{FB} models in Table \ref{tab:fiducial} with a series of smaller $a_{\rm supp,max}$.  We find that the results (i.e., $Z_{\rm final}$) do not change even with $a_{\rm supp,max} = 0.001$ cm, indicating that the feedback effect-assisted efficient coagulation is able to retain the supplied dust mass regardless of the size distribution of the dust supply.

We find that solid accumulation is overall faster and more significant with lower $\alpha$, higher $u_{\rm f}$, and larger $Z_{\rm supp}$, consistent with our findings from the fiducial models.  We consider such parameter combinations the optimal disc conditions.  With $Z_{\rm supp}=0.01$, only the run with the most optimal disc conditions (i.e., $\alpha = 1$e-$3$ and $u_{\rm f}=1000$ cm s$^{-1}$) yields significant dust accumulation within the simulation time, where $Z_{\rm final}=6.95$.  The value of $Z_{\rm final}$ declines immediately with increasing $\alpha$ and decreasing $u_{\rm f}$, both of which increase/decrease the possibility of fragmentation/coagulation and hence reduce the effectiveness of feedback effect.  Furthermore, the feedback effect is assumed to reduce dust diffusion (via $\alpha_{\rm FB}$; see Section \ref{subsubsec:feedback}) and reduce relative turbulent velocities between particles, which is equivalent to a parallel reduction in $u_{\rm f}$.  It thus appears that $Z_{\rm final}$ has a slightly stronger dependency on $\alpha$ than on $u_{\rm f}$.

With $Z_{\rm supp}=0.05$, the number of cases that retain substantial dust mass increases and $Z_{\rm final}$ in the run with the optimal disc conditions is roughly 5 times higher than that with $Z_{\rm supp}=0.01$ ($37.40$ versus $6.95$).  The dependence of $Z_{\rm final}$ on $\alpha$ and $u_{\rm f}$ has a similar general trend as observed in cases with lower $Z_{\rm supp}$.  When $u_{\rm f}=1000$ cm s$^{-1}$, $Z_{\rm final} = 50.03$ in the run with $\alpha = 1.78$e-$3$ and is higher than that ($37.40$) in the run with $\alpha = 1$e-$3$.  The reason is simply that $\Sigma_{\rm g}$ in our disc model decreases with $\alpha$.  Thus, these runs actually accumulate comparable amount of dust mass in order of magnitudes at the end of simulation ($\Sigma_{\rm d} \approx 54518$ g cm$^{-2}$ for $\alpha = 1$e-$3$ and $\Sigma_{\rm d} \approx 45825$ g cm$^{-2}$ for $\alpha = 1.78$e-$3$) with the maximized feedback effect.

Figure \ref{fig:Zfinal_capFB_noMT} also shows that the prediction of $Z_{\rm final}$ from our toy model (Equation \ref{eq:Z_final}) applies to most of the cases that end with a quasi-steady state without significant dust growth or accumulation.  With $Z_{\rm supp}=0.01$, $Z_{\rm final} \sim 3.3$ per cent, similar to that in Model \texttt{B1\_MT} (see Section \ref{subsubsec:mt_only}).  With $Z_{\rm supp}=0.05$, $Z_{\rm final} \sim 16$ per cent, again five times higher.

Figures \ref{fig:Zfinal_capFB_MT} and \ref{fig:Zfinal_capFB_MT_Z5} present our survey results on the \texttt{MT\_FB} models.   As discussed in Sections \ref{subsubsec:mt_only} and \ref{subsubsec:coopetition}, significant dust accumulation only happens with runaway growth, which requires efficient grain growth towards the size regime where $\Delta u \lesssim u_{\rm f}$ (i.e., $>10^{3}$ cm).  Otherwise, mass transfer would compete with feedback effect for solid mass and damp the latter mechanism, leading to little or none dust retention.  Consequently, the dust evolution results for $a_{\rm supp,max} \leqslant 300$ cm are almost identical and are represented by models with $a_{\rm supp,max} = 300$ cm in Figures \ref{fig:Zfinal_capFB_MT} and \ref{fig:Zfinal_capFB_MT_Z5}. 

Fast and robust dust accumulation is possible with $a_{\rm supp} = 1000$ cm and with the optimal (or near-optimal) disc conditions.  Since the onset of the runaway growth depends on whether the self relative velocities of the largest unmodulated particles available in the size distribution drop below $u_{\rm f}$, the number of cases with $Z_{\rm final} \gg 1$ drops dramatically with decreasing $u_{\rm f}$ and depend slightly less on $\alpha$.  Once significant breakthrough happens, the dust evolution is dominated by mass transfer, which transports almost all the supplied dust mass to the very large particles produced by runaway growth.  Thus, the total accumulated solid masses in such cases are comparable given enough evolution time, which explains the findings that $Z_{\rm final}$ is roughly inversely proportional to $\Sigma_{\rm g}$ for a given $Z_{\rm supp}$ and is hence larger with increasing $\alpha$.  The second consequence is that the transition in $Z_{\rm final}$ between cases with and without significant dust accumulation is more abrupt than that observed in the \texttt{FB} models in Figure \ref{fig:Zfinal_capFB_noMT}.

In most of the \texttt{MT\_FB} models that fail to retain substantial dust mass, $Z_{\rm final}$ is again in good agreement with our simple prediction (Equation \ref{eq:Z_final}), that is, $\sim 3.3$ and $\sim 16$ per cent for $Z_{\rm supp}=0.01$ and $Z_{\rm supp}=0.05$, respectively.

%%%%%%%%%%%%%%%%%%%%%%%%%%%%%%%%%%%%%%%%%%%%%%%%%%%%%%%%%%%%%%%%%%%%%%%%%%%%%%%
\section{Summary and Discussions}
\label{sec:final}

% what we did
We study the dynamic local evolution of solids near the inner boundary of Class II PPDs.  At the assumed disc evolutionary stage, the host T Tauri stars have acquired nearly all their asymptotic masses and the accretion rate is reduced to the extent where the magnetospheric truncation radius expands and the gas temperature nearby decreases, such that refractory grains are preserved in a condensed state.

% why it is interesting
The dynamic dust evolution at such late stages of disc evolution is of great astrophysical interest, since the naturally formed global pressure maximum near the disc inner edge can trap dust grains and potentially produce planet building materials and even planets.  This process offers a pathway to form the abundant close-in super-Earth and sub-Neptune (also known as Kepler planets).  Moreover, these planets are likely statistically disconnected from their host metallicity that is largely determined beforehand.  Therefore, the proposed dust growth/accumulation mechanism in this work may naturally solve the conundrum that Kepler planets are insensitive to stellar metallicity.
% RL: one may argue that $Z_{\rm supp}$ may still be connected to stellar metallicity, we may argue that $Z_{\rm supp}$ is likely regulated by other planetesimal formation processes and is also disconnected from stellar metallicity.

\subsection{Synopsis of Basic Results}
% more on our simulations
In this work, we first construct a series of radiative disc models based on the requirement that the gas temperature around the dust accumulation disc radius $R_{\rm accu}\approx R_{T}$ lies between $\approx$1000 and $\approx$2000 K.  We then carry out a set of local dust evolution simulations at $R_{\rm accu}$ with our newly-developed implicit coagulation-fragmentation code, \texttt{Rubble}.  Our numerical model evolves the dust size distribution by solving the Smoluchowski equation with a comprehensive particle collision model (described in Section \ref{subsec:base_model}; see also Figure \ref{fig:scenarios}).  This work for the first time incorporates dust feedback in a local dust evolution model, where heavy dust-loading damps particle diffusion in gas (see Section \ref{subsubsec:feedback}).  In addition, \texttt{Rubble} dynamically evolves the total dust surface density by taking into account a prescribed dust supply carried in by disc accretion and the dust loss in small grains carried away by accretion funnels onto the protostar (described in Section \ref{subsec:dust_evo}).  

% refer to result section
We are particularly interested in whether or not and how much dust can be retained at $R_{\rm accu}$, which turns out to be heavily dependent on the dominant physical processes in dust evolution, and the combination of key parameters of interest ($\alpha$, $u_{\rm f}$, $a_{\rm supp,max}$, and $Z_{\rm supp}$; see Section \ref{subsec:setup}).  Since accretion funnels constantly drain solids below a certain size ($\sim 10^{-2}$ cm), only large particles can survive in the long run.  Consequently, processes that favour a prolific production of small grains would deplete the inventory of planet-building materials, while processes that lead to efficient dust growth may accumulate significant dust mass.  We thus specifically focus and experiment on the latter processes, including mass transfer and feedback effects (see Section \ref{subsec:fiducial_cases}).  To further identify the threshold for significant dust accumulation, we survey each key parameter within a physically motivated range (see Section \ref{subsec:survey_results}).

% summarize results/scenarios
We find that our simulation results can be categorized into the following three scenarios:
\begin{enumerate}
  \item \textbf{Equilibrium Scenario} -- a quasi-steady state without significant dust growth or accumulation, where the dust supply rate balances the dust loss rate (i.e., the dust removal efficiency $\dot{\Sigma}_{\rm d,out}/\dot{\Sigma}_{\rm d,in}$ is unity);
  \item \textbf{Feedback + GI Scenario} -- efficient retention and accumulation of cm-sized grains that serve as a mass reservoir, where the dust surface density $\Sigma_{\rm d}$ monotonically increases with low but non-zero dust removal efficiency and eventually results in planetesimal formation via gravitational instability (GI);
  \item \textbf{Breakthrough Scenario} -- effective break-through growth across the fragmentation barrier that leads to runaway buildup of larger and larger particles toward planetesimals, where $\Sigma_d$ surges due to the nearly zero dust removal efficiency. 
\end{enumerate}

\subsection{Implications}
\label{subsec:implications}

In the Equilibrium Scenario, stars efficiently accrete all the heavy elements to their proximity.  The incorporation of a rich population of small grains in the magnetic funnels may provide an effective opacity source in the stellar magnetosphere, which could periodically obscure the observed stellar flux.  To roughly estimate the optical depth of accretion funnels, we assume that they launch axisymmetrically from $R_{\rm T}$ over a radial extent $\textnormal{d} R$.  Near the launching point, the mean funnel gas density is then $\rho_{\rm f} \sim {\dot M}/(2 \pi R_{\rm T} \textnormal{d} R V_{\rm z})$ and the optical depth across the funnel is
\begin{equation}
  \begin{split}
    \tau_{\rm f} \sim \kappa_{\rm f} \rho_{\rm f} \textnormal{d} R \sim 7.54 &\left(\frac{\dot{M}}{3\times 10^{-9} M_\odot \ \text{yr}^{-1}}\right) \left(\frac{\kappa_{\rm f}}{30\ \text{cm}^{2}\ \text{g}^{-1}}\right) \\
    & \left(\frac{R_{\rm T}}{0.08\ \text{au}} \right)^{-1} \left(\frac{V_{\rm z}}{1\ \text{km}\ \text{s}^{-1}} \right)^{-1},
  \end{split}
\end{equation}
where the flow speed $V_{\rm z}$ is approximated as $c_{\rm s}$ and the funnel opacity $\kappa_{\rm f}$ is estimated to be a few times that of the solar value \citep{BellLin1994}.  The accretion funnels are thus optically thick upon launching.  However, they likely become optically thin when approaching the magnetic poles of the host star, where $V_{\rm z}$ is accelerated to the free-fall velocity and $\kappa_{\rm f}$ substantially decreases inside the grain sublimation radius such that $\tau_{\rm f} \ll 1$.  It is also possible for dust grains to survive for a while in the magnetosphere and contribute opacity if the sublimation timescale is comparable to the free-fall timescale \citep{Nagel2020}.

The initially opaque but later transparent accretion funnels, along with the non-axisymmetric and variable disc accretion, may account for some commonly but not universally observed ``dippers'' in the light curves of T Tauri stars with strong magnetic fields and relatively active circumstellar discs \citep[][etc.]{Bouvier1999, Cody2014, Ansdell2016a, Roggero2021}.  Some dimming events appear to be quasi-periodic, though more extended follow-up observations are needed to establish their nature.  Future detailed analyses of their light curves and thorough comparisons to magnetospheric accretion models \citep[e.g.,][]{Mcginnis2015, Bodman2017} are required to constrain the evolution of dust inside the magnetic funnel as well as variabilities of magnetosphere-disc interactions.

% second scenario, explain connections to the GI
In the Feedback + GI Scenario, the efficient accumulation of dust surface density may trigger multiple planetesimal formation mechanisms, such as direct gravitational collapse \citep[or direct GI; e.g.][]{Goldreich1973, GaraudLin2007}, the streaming instability \citep[SI; e.g.][]{Youdin2005b,Johansen2007,Simon2017,Li2019,Carrera2021}, vortices trapping \citep{Johansen2007}, etc.  Our models do not mimic planetesimal formation by removing dust mass under certain conditions as seen in some previous works \citep[][etc.]{Drazkowska2014b, Stammler2019} because such a treatment introduces extra dependencies and uncertainties.  Below we estimate the solid abundance needed to trigger the direct GI and the SI.

At the center of the pressure bump, direct GI take place once the midplane dust density exceeds the Roche density $\rho_{\rm Ro} = 9M_\star / (4\pi R_{\rm accu}^3) \sim 4.5\times 10^{-4}\ \text{g cm}^{-3}$.  The corresponding dust-to-gas surface density ratio is roughly
\begin{equation}
  Z_{\rm GI} \sim \frac{\sqrt{2\pi} \rho_{\rm Ro} H_{\rm d}}{\Sigma_{\rm g}} \sim 184 \left(\frac{H_{\rm d}/H}{0.005}\right) \left(\frac{\alpha}{0.001}\right)^{7/10},
\end{equation}
where we estimate $Z_{\rm GI}$ using the dust scale height of the dominant species, i.e., a few cm-sized grains around the peak of the final size distribution.  The largest $Z_{\rm final}$ observed in the \texttt{FB} models is a few times lower than the estimated $Z_{\rm GI}$, suggesting that the onset of GI may require a longer evolution.  That said, taking into account the size-dependent velocity dispersion \citep{Volkov2000}, the non-axisymmetric distribution of dust mass, or alleviating the numerical limitation of feedback effect may make it easier for the GI to happen, which merits further investigation.

% SI on the edge of pressure bump
As dust accumulates across the pressure bump, planetesimals may also form astride the bump via the SI .  We again assume that one dust species dominates the dynamics and follow the strong clumping criteria in \citet[][see their Section 3.3 and Equations 11 and 14]{Li2021} to estimate the required dust-to-gas surface density ratio
\begin{equation}
  \begin{split}
    &Z_{\rm SI} \simeq \epsilon_{\rm crit}(\uptau_{\rm s}) \sqrt{\left(\frac{\Pi}{5}\right)^2 + \frac{\alpha}{\alpha + \uptau_{\rm s}}} \sim 0.15, \\
    &\text{where}\ \Pi \equiv -\frac{c_{\rm s}}{2 \Omega_{\rm K} R} \frac{\partial \ln \rho_{\rm rad}}{\partial \ln R}
  \end{split}
\end{equation}
represents the strength of the global radial pressure gradient and we adopt $\Pi = 0.26$, $\alpha=0.01$, and $\uptau_{\rm s} = 0.05$, which corresponds to a few cm-sized dust.  Our simple order of magnitude estimation thus indicates that it may be easier to trigger the SI than the direct GI to produce planetesimals.  Consequently, the SI may convert part of the supplied solids to planetesimals in the wings of the pressure bump and slow down the dust accumulation at the center.  Nevertheless, the value of $Z_{\rm SI}$ for the SI with a dust size distribution is likely of order a few higher but is poorly constrained so far.  Future extensive studies are needed to better understand and compare different collective mechanisms for planetesimal formation around $R_{\rm accu}$. 

% connect to dipper as well
The Feedback + GI Scenario implicitly assumes that mass transfer is ineffective due to, for example, a much higher mass ratio ($\gg 50$) required between the colliding particles.  This assumption actually frees our model from one of the parameter dependencies and make cm-sized dust grains the key population for mass accumulation.  In this way, the elevated $\Sigma_{\rm d}$ intensifies collisional frequencies, resulting in more small grains subject to funnel removal.  The dust removal efficiency thus remains low but non-zero in the rapid accumulation stages (see Figure \ref{fig:fiducial_dust_removal}) and then gradually increases towards unity as the dust-loading approaches the capacity of the mass reservoir.  However, planetesimal formation may kick in regularly, consume some dust, and reactivate rapid solid accumulation, causing the downturns in dust removal efficiency.  This process may also regulate the opacity of the magnetic funnels, leading to aperiodic or quasi-periodic dimming events.  Even if the full mass reservoir does not trigger planetesimal formation, the heavy dust-loading may increase the dust disc thickness near $R_{\rm accu}$ (not modelled in this work) and cast long shadows onto the outer disc.  Light modulation due to such shadows may account for the variability in the multi-epoch imaging of scattered light around some resolved PPDs \citep{Stolker2017, Pinilla2018b}.

% third scenario
In the Breakthrough Scenario, planetesimals emerge directly from grain growth.  The efficient sweep-up growth of larger particles through mass transfer almost depletes small grains and leads to severe reduction in the dust removal efficiency.  Thus, the opacity ($\kappa_{\rm f}$) and optical depth ($\tau_{\rm f}$) along the magnetic funnels and the number of possible occultation events drop significantly.  Indeed, dippers are only observed in a fraction of T Tauri stars with an occurrence rate of $\sim 30$ per cent \citep{Cody2014,Hedges2018,Cody2018}.  Future statistical studies on the relationship between disc properties and the dipper occurrence rate may enable us further distinguish and constrain planetesimal formation scenarios. 

% what happens after planetesimal formation
Once planetesimals form around $R_{\rm accu}$, they continue to grow towards super-Earths via pebble accretion \citep{Ormel2010, Ormel2017}.  Their growth may be quenched when they reach the pebble isolation mass \citep{Lambrecht2014, Bitsch2018}, albeit collision among the accumulating pebbles near the planets' tidal barrier may enable small grains to bypass this dust dam \citep{ChenYX2020}.  The enhanced opacity and the relatively high entropy in the proximity of super-Earths limit their potential to accrete non-negligible atmosphere through Kelvin-Helmholtz contraction \citep[e.g.][]{Pollack1996, PisoYoudin2014, LeeChiang2014, LeeChiang2015, Ali-Dib2020}.  The formation efficiency of these subsequent processes calls for future research.

Furthermore, we emphasize that the dust accumulation radius $R_{\rm accu}$ does not directly reflect the final orbital configuration of close-in planets.  Many mechanisms have been proposed to evolve and alter their orbits \citep[][etc.]{Dobbsdixon2004, Mardling2007, Kley2012} and explain their occurrence rate \citep[e.g.,][]{Leechiang2017}.  For instance, an external non-coplanar massive planetary or stellar companion may excite the eccentricity of close-in planets via secular interaction or Kozai-Lidov effect.  Such eccentric orbits may further be shrunk and circularized by tidal effects \citep[see the review in][]{Naoz2016}.  Additionally, close-in planets orbiting around protostars with strong magnetic field may go through orbital evolution in either direction due to the Alfv\'en drag or unipolar induction \citep{Lainelin2012}.  Future parameter surveys are needed to make connections between the dust accumulation radius and the statistical orbital distribution of close-in planets.

\subsection{Limitations}
\label{subsec:limitations}

% limitation of this work
Finally, we note that our results are subject to several limitations.  Our study evolves dust distributions locally in a static axisymmetric radiative disc profile.  However, the gas disc evolves dynamically \citep[e.g.,][see their Section 5.3]{ChenYX2020b} and may be non-axisymmetric to some extent.  For example, we adopt a default opacity of $\kappa=1$cm$^2$ g$^{-1}$ to construct the disc temperature profile based on empirical opacity tables \citep{BellLin1994}, whereas real opacity depends on solid abundance.  As dust accumulates, the opacity roughly scales linearly with the amount of sub-mm-sized grains \citep{Ormel2014}.  Thus, the gas temperature $T_{\rm rad} \propto \kappa^{1/5} \approxprop Z_{\rm sub-mm}$, that is, $T_{\rm rad}$ doubles when $Z_{\rm sub-mm}$ increases $\sim 40$ times.  That said, accretion funnels keep removing sub-mm-sized dust and solid mass accumulates in either cm-sized grains or planetesimals, suggesting that $Z_{\rm sub-mm}$ does not necessarily increase with $Z$.  Nonetheless, future self-consistent calculations are needed to model the mutual influence between dust and opacity/gas temperature at $R_{\rm accu}$ \citep{ChenYX2020, Savvidou2020}.

Admittedly, our dust removal prescription is idealized in many aspects and therefore bears some uncertainties.  We first note that our prescription differs with those photoevaporative disc wind models applicable to the outer regions of PPDs \citep{HutchisonClarke2021, BoothClarke2021}.  The driving mechanism for funnel flows is the magneto-centrifugal force that dominates the pressure gradient in the low-density disc atmosphere above $\sim H$ \citep{Blandford1982, Wardle1993}.  Consequently, it is likely that larger grains may be entrained by the funnel flows than the photoevaporative winds.  In Section \ref{subsec:funnel_flows}, we assume that the launching height of funnel flows to be about one gas scale height ($H_{\rm s} \sim H$) and then estimate relevant quantities for dust removal based on a Gaussian profile for the dust vertical distribution (e.g., $r_{\rm crit,D}$ and $f_H$; see Equations \ref{eq:r_d_critD} and \ref{eq:f_H}).  In our models, $f_H$ slightly depends on $\alpha$ and is roughly $0.3$ for most dust smaller than $r_{\rm crit,D}$.  Models with different assumptions (e.g., different vertical profiles) may lead to a larger/smaller $f_H$ and hence a factor of a few faster/slower dust removal.  Such moderate changes in $f_H$ are expected to have more direct impacts on $Z_{\rm final}$ (see Equation \ref{eq:Z_final}) in the Equilibrium Scenario, whereas in the Feedback + GI and Breakthrough Scenarios, dust accumulation eventually dominates over dust removal.  Nevertheless, further detailed modeling of dusty disc-star interactions via magnetic torques is needed to refine all these assumptions and improve our results.

Our local model also neglects the evolution of the global disc, whereas $\alpha$, $a_{\rm supp,max}$, and $Z_{\rm supp}$ are likely time-dependent and have radial profiles.  Modelling dust evolution with extra dimensions, such as the radial dimension and even the vertical dimension, is crucial for better understanding realistic dust growth and accumulation \citep{Drazkowska2013}.  Moreover, $a_{\rm supp,max}$ and $Z_{\rm supp}$ are likely regulated by pressure bumps and planetesimal formation processes (i.e., the SI) outside $R_{\rm accu}$.  For instance, a sharp transition in the radial profile of $\alpha$ at the DZIB may naturally form a pressure bump that trap dust grains and produce planetesimals \citep{Chatterjee2014, HuXiao2016}.  However, dust trapping in local pressure maxima may be lossy depending on the bump amplitudes \citep{Pinilla2020, Carrera2021} and pebbles could still diffuse through these maximum and drift inwards \citep{LiYaPing2019}.  These pebbles would be then captured by the global pressure maximum at $R_{\rm accu}$.  Furthermore, it is possible that $R_{\rm accu}$ is very close the location of the DZIB in realistic discs, considerably complicating the modelling of the inner disc.  Future works are required to simulate the adjacent global \& local pressure bumps self-consistently and comprehend the dust evolution therein.
\footnote{Previous works have examined the close relation of the sublimation front and the DZIB \citep{Flock2016,Ueda2017,Ueda2019,Flock2019} using radiative transfer and hydrodynamical simulations. Their results apply for discs with very low accretion rates $\dot{M}\sim 10^{-10} M_{\odot}$ yr$^{-1}$ when stellar irradiation dominates over viscous heating.  How the picture changes with an active disc region remains to be studied.}

In addition, this work only explore one set of stellar parameters (see Section \ref{subsec:setup} and Table \ref{tab:paras}), which fundamentally determine the disc profile ($T_{\rm rad}$, $\dot{M}$, etc.) and the maximum grain size that can be lifted by funnel flows.  Studying the stellar dependency of the dust evolution requires traversing a much larger parameter space.  We thus leave it to future surveys.

%%%%%%%%%%%%%%%%%%%%%%%%%%%%%%%%%%%%%%%%%%%%%%%%%%%%%%%%%%%%%%%%%%%%%%%%%%%%%%%%
\section*{Acknowledgements}

We thank Chris Ormel, Til Birnstiel, Andrew Youdin, Kaitlin Kratter, Dong Lai, Saul Rappaport, Gibor Basri, and Lee Hartmann for useful discussions.  

\section*{Data Availability}

The data in this article are available from the corresponding author on reasonable request. 

%%%%%%%%%%%%%%%%%%%%%%%%%%%%%%%%%%%%%%%%%%%%%%%%%%


\bibliographystyle{mnras}
\bibliography{refs}

%%%%%%%%%%%%%%%%%%%%%%%%%%%%%%%%%%%%%%%%%%%%%%%%%%

%%%%%%%%%%%%%%%%% APPENDICES %%%%%%%%%%%%%%%%%%%%%

\appendix
%%%%%%%%%%%%%%%%%%%%%%%%%%%%%%%%%%%%%%%%%%%%%%%%%%%%%%%%%%%%%%%%%%%%%%%%%%%%%%%%
\section{Tests of the Dust Evolution Code}
\label{app:tests}

In this section, we first test our numerical model against pure coagulation cases with well defined analytical predictions.  Sections \ref{appsubsec:frag_test} and \ref{appsubsec:more_tests} then present test cases with additional physical ingredients against empirical results in previous studies.

\subsection{Simple Coagulation Kernels with Known Analytical Solutions}
\label{appsubsec:sim_kernel}

\begin{figure*}
  \centering
  \includegraphics[width=0.32\linewidth]{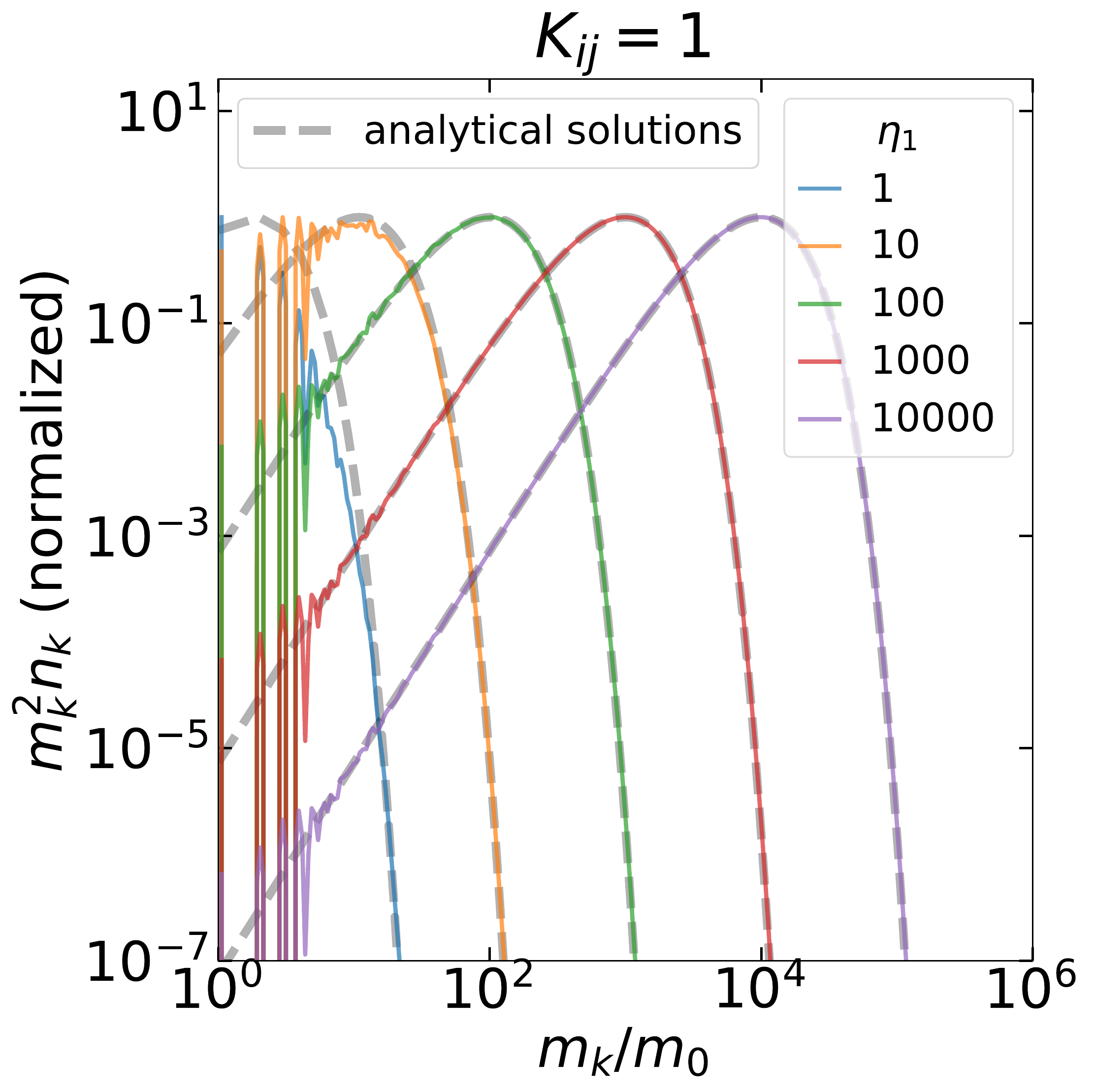}
  \includegraphics[width=0.32\linewidth]{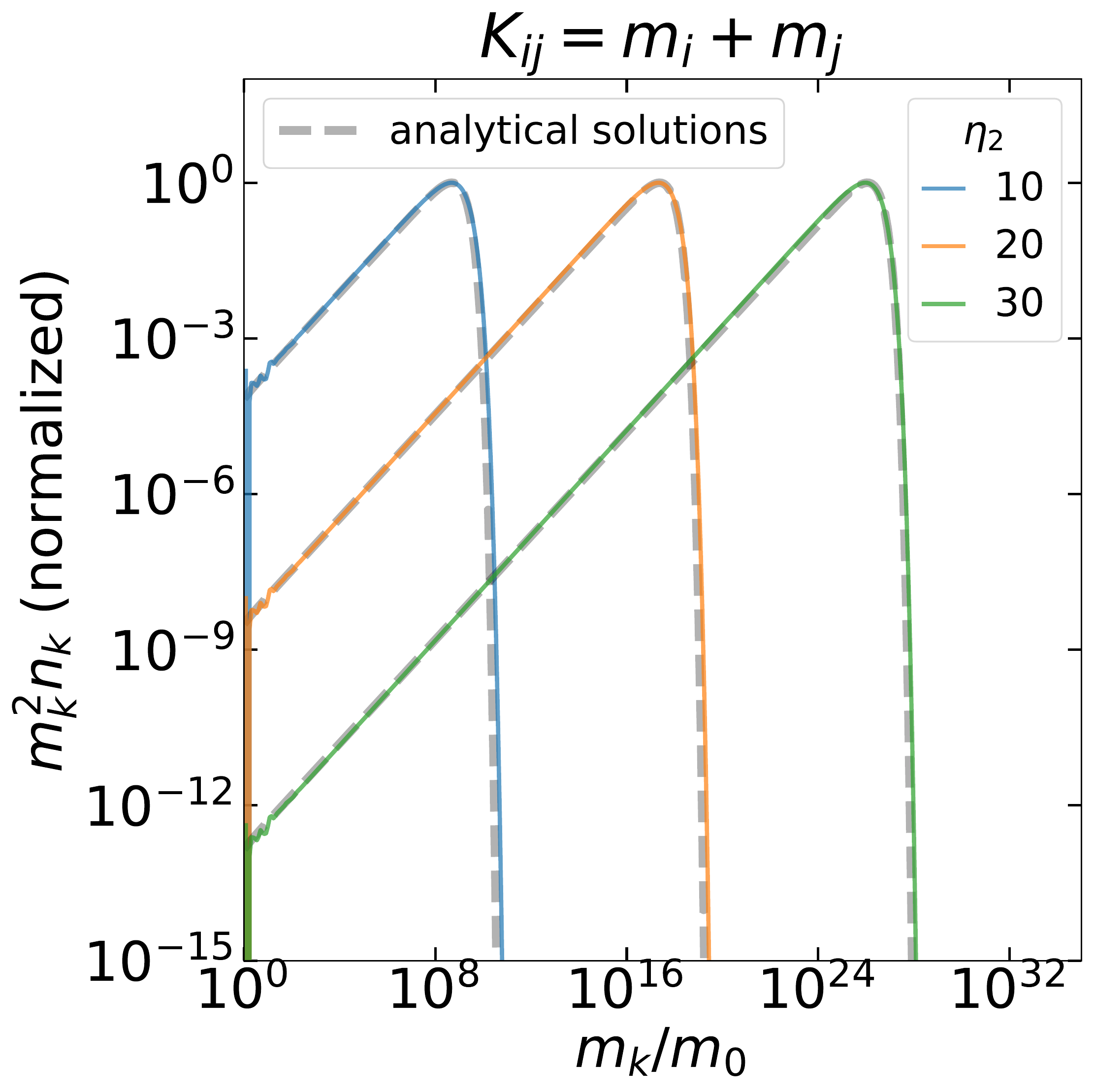}
  \includegraphics[width=0.32\linewidth]{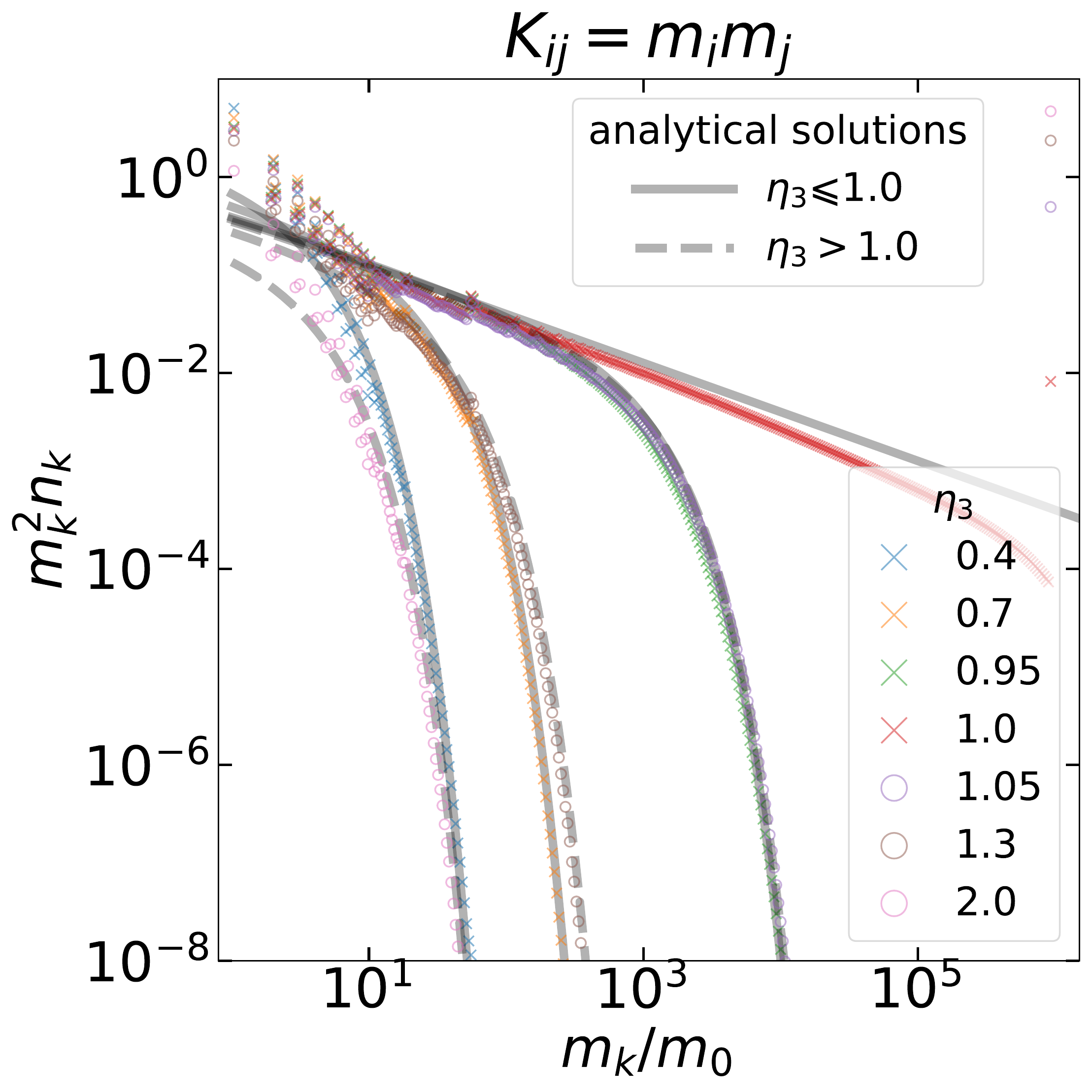}
  \caption{Evolution of the mass distribution for a constant coagulation kernel  ({\it left}), $K_{ij} = 1$, for a sum kernel ({\it middle}), $K_{ij} = (m_i + m_j)$, and for a product kernel ({\it right}), $K_{ij} = m_i m_j$.  The simulation results ({\it colour curves and markers}) are in good agreement with the analytical solutions ({\it grey curves}), where $\eta_1$, $\eta_2$, and $\eta_3$ are the dimensionless timescales related to these kernels (see Table \ref{tab:sln4coag}), respectively.  The discontinuities of the simulation results on the small mass end (especially in the left and right panels) are due to the discrete nature of the numerical mass grid. \label{fig:analyK}}
\end{figure*}

With the fragmentation part neglected, the Smoluchowski equation (Equation \ref{eq:Seq}) has analytical solutions for three simple coagulation kernels, namely the constant kernel, the sum kernel, and the product kernel (\citealt{Wetherill1990, Kenyon1998, Ormel2008}; see also Appendix A4 in \citet{Kenyon1998} and Section 3.2 in \citet{Ormel2008} for more details ).  Table \ref{tab:sln4coag} summarizes the predicted mass distributions for these kernels that evolves from an initially monodisperse particle distribution as a function of time and the parameter choices in our numerical tests.  Figure \ref{fig:analyK} shows that our numerical solutions are able to closely follow the analytic model and produce good agreement.  In the product kernel test, we enable the coagulation between the right ghost mass bin and other mass bins in order to let the runaway growth continue to consume mass after $\eta_3 > 1$.

\begin{table*}
  \caption{Analytical solutions for the Smoluchowski Equation with Simple Kernels \label{tab:sln4coag}}
  \begin{tabular}{c|c|c|c|c}
  \hline
  kernel name &
  \makecell{kernel {$K_{ij}$}} &
  \makecell{dimensionless\\ time unit} &
  solutions $n_k(n_0, \eta)$ $^{*}$ &
  tested with \\  
  \hline\hline
  \rule{0pt}{0.75cm}
  Constant & $\alpha_{\rm c}$            & $\eta_1 = \alpha_{\rm c} n_0 t$ & 
  $ \displaystyle \begin{aligned}
      &n_k = n_0 f^2 (1 - f)^{k-1}, \\
      &\text{where}\ f = 1/(1+\eta_1/2)
  \end{aligned} $ &
  \makecell{$(\alpha_{\rm c}, n_0) = (1, 1)$ \\254 logarithmic mass bins\\ for $k=1,\cdots,10^{6}$}
  \\[1em]
  \hline
  \rule{0pt}{0.75cm}
  Sum      & $\beta_{\rm c} (m_i + m_j)$ & $\eta_2 = \beta_{\rm c} n_0 t$  &
  $\displaystyle \begin{aligned}
    &n_k = n_0 \frac{k^{k-1}}{k!}f(1-f)^{k-1} \exp[-k(1-f)], \\
    &\text{where}\ f = \exp(-\eta_2)
  \end{aligned}$ &
  \makecell{$(\beta_{\rm c}, n_0) = (1, 1)$\\374 logarithmic mass bins\\ for $k=1,\cdots, 5.26\times10^{28}$}
  \\[2em]
  \hline
  \rule{0pt}{0.75cm}
  Product  & $\gamma_{\rm c} m_i m_j$    & $\eta_3 = \gamma_{\rm c} n_0 t$ &
  $\displaystyle n_0 \frac{(2k)^{k-1}}{k!k}(\eta_3/2)^{k-1} \exp(-k\eta_3)$ &
  \makecell{$(\gamma_{\rm c}, n_0) = (1, 1)$\\ 374 logarithmic mass bins\\ for $k=1,\cdots,10^{6}$}
  \\[1em]
  \hline
  \end{tabular} \\
  \begin{flushleft}
    {\large N}OTE --- $^{*}$ The solutions shown in this table are the predicted mass distributions evolved from an initially monodisperse particle distribution, i.e., there are only $n_0$ particles with the same mass $m_0$ at the beginning. In the solutions, $n_k$ denotes the number of particles with mass $m_k = k m_0$.
  \end{flushleft}
\end{table*}

\subsection{Coagulation-Fragmentation in Disc Environments}
\label{appsubsec:frag_test}

\begin{figure*}
  \centering
  \includegraphics[width=0.8\linewidth]{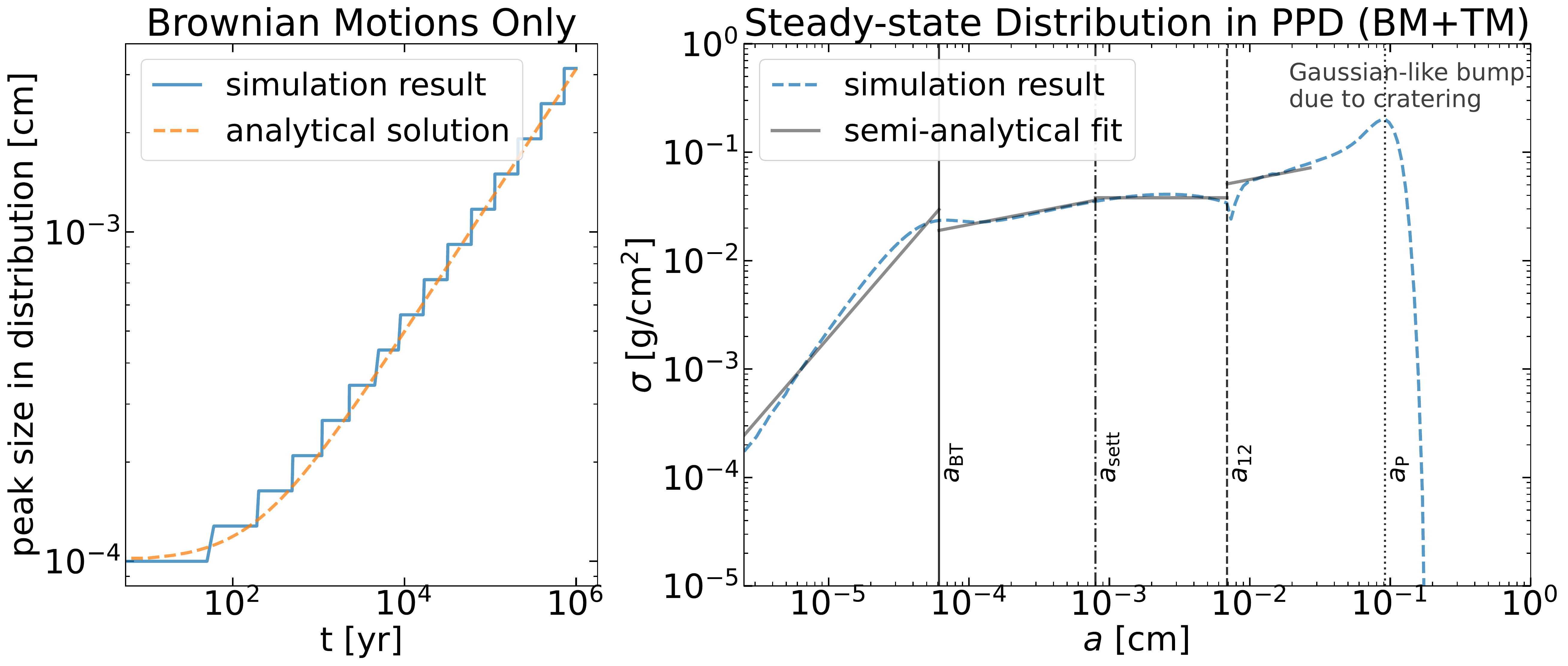}
  \caption{Comparisons between numerical and semi-analytical results for growth of a monodisperse particle distribution in a typical disc environment (see Section \ref{appsubsec:frag_test} for details), where dust evolves with only the Brownian motions between the same sized particles (\textit{left}) and with both Brownian motions and turbulent relative motions between all particles (\textit{right}).  For the first test (see \citealt{Birnstiel2010} for derivations), the left panel shows the simulated peak size of dust distribution (\textit{blue solid}) strictly follows the predicted values (\textit{orange dashed}) .  For the second test, the right panel shows the slopes of the simulated distribution (\textit{blue dashed}) can be closely fitted by the analytical predictions (\textit{grey solid}) in different regimes, namely the Brownian motion regime ($<a_{\rm BT}$), the turbulent regime ($a_{\rm BT}$ -- $a_{\rm sett}$), the turbulent regime with settling effects ($a_{\rm sett}$ -- $a_{\rm 12}$), and the strong turbulent regime ($a_{\rm 12}$ -- $a_{\rm P}$). These regimes are separated by characteristic dust sizes (see \citet{Birnstiel2011} for derivations and the fitting recipe).  The small bump near the peak position of the distribution is due to the cratering effects.} \label{fig:BM_steady_state}
\end{figure*}

\begin{figure*}
  \centering
  \includegraphics[width=0.8\linewidth]{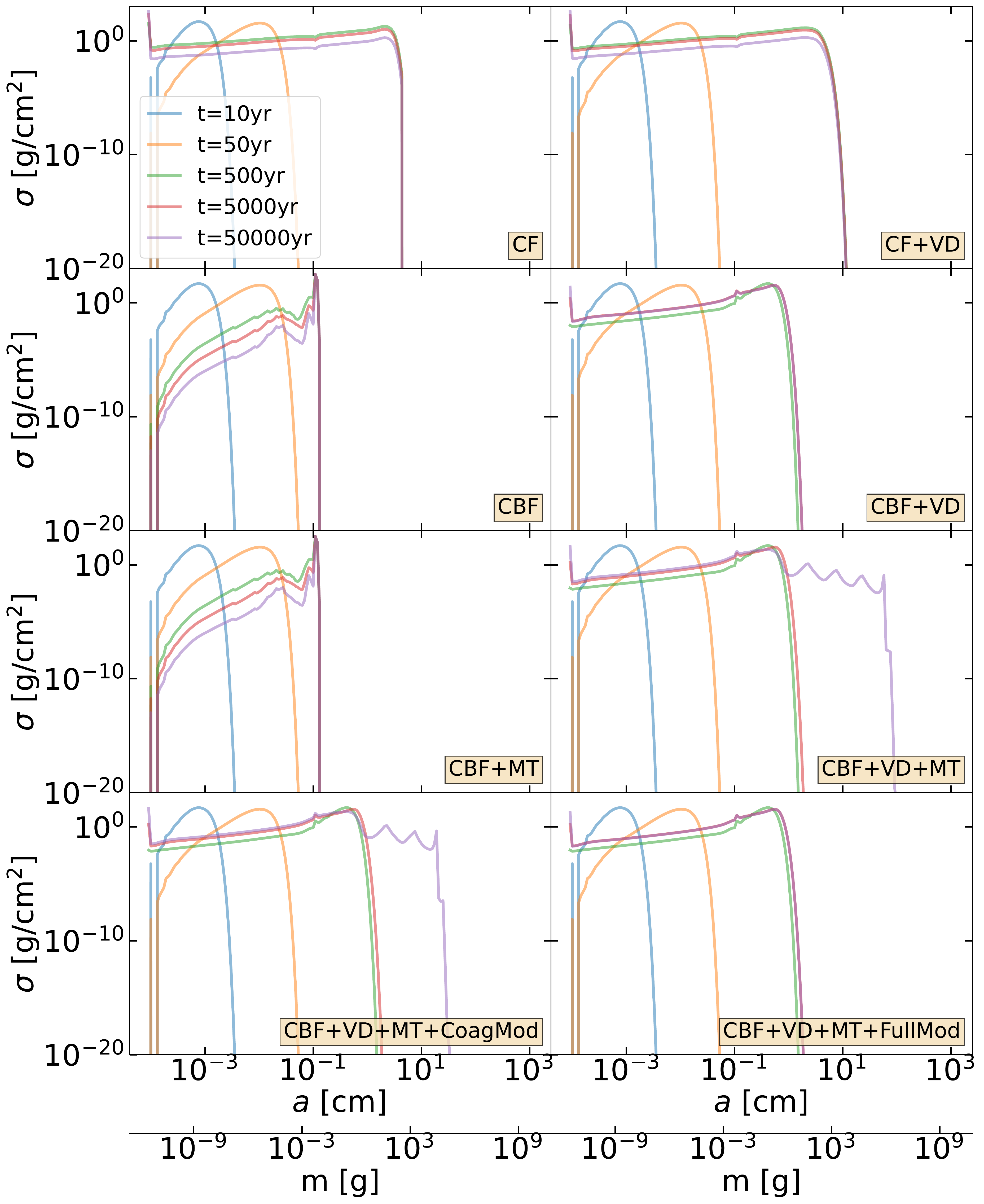}
  \caption{Comparison on the size distribution evolution over $5\times10^4$ years for models that include different physical ingredients (\textit{labelled} in each panel) in our collision treatment, where ``\textsf{CF}'' means ``coagulation and fragmentation'', ``\textsf{VD}'' represents ``velocity distribution'', ``\textsf{CBF}'' stands for ``coagulation, bouncing, and fragmentation'', ``\textsf{MT}'' denotes ``mass transfer'', ``\textsf{CoagMod}'' indicates ``modulation on coagulation kernel only'', and ``\textsf{FullMod}'' means ``modulation on both the coagulation kernel and fragmentation kernel''.  No mass supply from outer reservoir is taken into account here. \label{fig:collision_model}}
\end{figure*}

To apply our numerical code in more realistic disc environments, we perform a test on the growth of a monodisperse dust distribution with only the Brownian motions between the same sized dust ($K_{i} \propto \sqrt{1/m_i}$).  For this kernel, there is no simple analytical solution on the evolution of the distribution, but the position of the peak of the size distribution may be approximated as \citep[see Equation B.5 in][]{Birnstiel2010}
\begin{equation}
  a(t) = \left(\frac{5}{2} \frac{\Sigma_{\rm g}}{\sqrt{2\pi}H}\frac{1}{\pi \rho_\bullet} \sqrt{\frac{12 k_{\rm B} T}{\rho_\bullet}} t + a_0^{5/2} \right)^{2/5}
\end{equation}
if the coagulation probability is unity, where $k_{\rm B}$ is the Boltzmann constant and $a_0$ is the size of the initial monodisperse dust.  Our test simulation adopts $T=196$ K, $\rho_\bullet=1.6$ g cm$^{-3}$, $\Sigma_{\rm g}=18$ g cm$^{-2}$, $\Sigma_{\rm g}=0.18$ g cm$^{-2}$, $H = 9.33\times10^{11}$ cm, and $\alpha=10^{-3}$. Figure \ref{fig:BM_steady_state} shows that the growth behaviour of such a dust distribution in our simulation is in a good agreement with the prediction.

To further test the fragmentation part of our numerical code, we follow \citet{Birnstiel2011} and conduct another test on the quasi-equilibrium dust distribution under typical disc conditions, with both the Brownian motions and turbulent relative motions between particles.  Our test adopts $T=50$ K, $\rho_\bullet=1.6$ g cm$^{-3}$, $\Sigma_{\rm g}=20$ g cm$^{-2}$, $\Sigma_{\rm g}=0.2$ g cm$^{-2}$, $H = 9.33\times10^{11}$ cm, $\alpha=10^{-4}$, $u_{\rm f}=100$ cm s$^{-1}$, and $\xi=1.83$.  Figure \ref{fig:BM_steady_state} shows the resulting steady-state distribution again evolved from a monodisperse dust distribution and demonstrates that our results reproduce the predicted slopes in all expected size regimes.

\subsection{Bouncing Barrier, Velocity Distribution, and Modulated Mass Transfer}
\label{appsubsec:more_tests}

In this section, we follow \citet{Windmark2012b, Drazkowska2014} to test the effects of including bouncing barrier, velocity distribution, mass transfer and modulation factors in our collision treatments.  We conduct a suite of simulations with varying configurations of physical ingredients and with the same disc parameters as indicated in Table 1 of \citet{Windmark2012b}.  Figure \ref{fig:collision_model} summarizes the time evolution of the dust size distributions for all of our simulations.

Comparing the final dust size distribution at $5\times 10^4$ yr in the top three rows of our Figure \ref{fig:collision_model} to those in Figure 2 of \citet{Windmark2012b}, we find that all cases yield similar results and manifest the effects of relevant physical ingredients, except the model \textsf{CBF+VD+MT}.  For this outlier, \citet{Windmark2012b} found that super-cm ``lucky particles'' quickly achieve runaway growth by sweeping up smaller particles via mass transfer.  Our simulation instead shows that the growth beyond cm is gradual and progresses through a ``step by step'' manner, as can be seen from the evenly-spaced chain of $\sigma(a)$ local maxima in Figure \ref{fig:collision_model}, where the step length is $\sim 50$ times in mass -- the characteristic ratio for mass transfer
\footnote{We conduct extra test simulations with different characteristic mass ratios for mass transfer and find that the step length scales with that mass ratio.  Moreover, we run extra resolution tests (up to $40$ grids per mass decade) and find that the step length does not depend on resolution.}
.  Such a slow growth is resulted from the high self relative velocity between $\sim 10$ cm-sized particles, which is of order $10^3$ cm s$^{-1}$ and is much larger than $u_{\rm f}$ ($100$ cm s$^{-1}$), indicating that the overall efficiency of sweep-up growth is modestly higher than that of fragmentation.  The exact reasons for the discrepancy between our result and that in \citet{Windmark2012b} remains elusive, we plan to implement a more complex collision treatment in \citet{Windmark2012} and perform further comparisons in future works.

The bottom row of Figure \ref{fig:collision_model} presents the effects of including modulation factor $f_{\rm mod}$ in our \textsf{CBF+VD+MT} model (see Section \ref{subsubsec:collisional_outcomes}) to limit the artificial growth of mass bins with unrealistic low number densities.  We first apply $f_{\rm mod}$ only on the coagulation kernel \citep{Drazkowska2014} and find that dust growth is modestly damped, indicating most solid mass is indeed transferred to larger particles via sweep-up growth.  It is therefore important to also apply $f_{\rm mod}$ on the fragmentation kernel.  With fully modulated kernels in the \textsf{CBF+VD+MT+FullMod} model, we find that growth beyond cm is completely suppressed, as the amount of large lucky particles are unrealistic low such that they should not interact with other mass bins.  This finding validates the use of $f_{\rm out}$ on the fragmentation kernel and further demonstrates the robustness of our dust evolution model.

%%%%%%%%%%%%%%%%%%%%%%%%%%%%%%%%%%%%%%%%%%%%%%%%%%

% Don't change these lines (only for MNRAS)
\bsp    % typesetting comment
\label{lastpage}

\end{document}